\documentclass[%
 reprint,
 amsmath,amssymb,
 aps,
 pre,
]{revtex4-2}

\usepackage{helvet,times}
\usepackage{titlesec}
\usepackage[labelfont=bf]{caption}
\usepackage[capitalize]{cleveref}
\crefdefaultlabelformat{\textbf{#2#1#3}} 
\crefname{figure}{\textbf{Figure}}{\textbf{Figures}} 
\Crefname{figure}{\textbf{Figure}}{\textbf{Figures}}

\newcommand{\pd}[2]{\frac{\partial #1}{\partial #2}} 
\newcommand{\pdd}[2]{\frac{\partial^2 #1}{\partial #2^2}} 

\newcommand{\SI}{{\bf SI}}
\newcommand{\fig}{\cref}
\newcommand{\eon}{\begin{equation}}
\newcommand{\eoff}{\end{equation}}

\newcommand{\sus}{\text{type A}}
\newcommand{\ifc}{\text{type B}}

\usepackage{xcolor}
\newcommand{\J}{\color{black}}

\usepackage{graphicx}
\usepackage{dcolumn}
\usepackage{bm}

\begin{document}

\preprint{APS/123-QED}

\title{Social network structure and the spread of complex contagions from a population genetics perspective}

\author{Julian Kates-Harbeck}
 \affiliation{Department of Physics,}
\author{Michael M. Desai}%
 \email{mdesai@oeb.harvard.edu}
\affiliation{%
Department of Organismic and Evolutionary Biology,\\ Harvard University, Cambridge MA~02138,~USA
}%

\date{\today}

\begin{abstract}
Ideas, behaviors, and opinions spread through social networks. If the probability of spreading to a new individual is a non-linear function of the fraction of the individuals' affected neighbors, such a spreading process becomes a ``complex contagion''. This non-linearity does not typically appear with physically spreading infections, but instead can emerge when the concept that is spreading is subject to game theoretical considerations (e.g. for choices of strategy or behavior) or psychological effects such as social reinforcement and other forms of peer influence (e.g. for ideas, preferences, or opinions). Here we study how the stochastic dynamics of such complex contagions are affected by the underlying network structure. Motivated by simulations of complex contagions on real social networks, we present a framework for analyzing the statistics of contagions with arbitrary non-linear adoption probabilities based on the mathematical tools of population genetics. {\J The central idea is to use an effective lower-dimensional diffusion process to approximate the statistics of the contagion. This leads to a tradeoff between the effects of "selection" (microscopic tendencies for an idea to spread or die out), random drift, and network structure. } Our framework illustrates intuitively several key properties of complex contagions: stronger community structure and network sparsity can significantly enhance the spread, while broad degree distributions dampen the effect of selection compared to random drift. Finally, we show that some structural features can exhibit critical values that demarcate regimes where global contagions become possible for networks of arbitrary size. Our results draw parallels between the competition of genes in a population and memes in a world of minds and ideas. Our tools provide insight into the spread of information, behaviors, and ideas via social influence, and highlight the role of macroscopic network structure in determining their fate.
\end{abstract}



\maketitle

\section{Introduction}
\subsection{Background}
Individuals on a social network are subject to influence by their neighbors, affecting their adoption of information \cite{zhou2015impact}, ideas \cite{pentland2014social}, and behaviors \cite{christakis2013social}. The likelihood that a given individual adopts a new idea depends on how many of her neighbors have adopted the idea already. For physically spreading infections, as encountered in traditional epidemiology \cite{keeling1999effects}, this dependence is typically linear and leads to a ``simple contagion''. By contrast, social reinforcement and other forms of peer influence \cite{weng2013virality,nematzadeh2014optimal}, as well as game theoretical considerations of behavior \cite{montanari2010spread}, can result in a non-linear dependence of an individual's likelihood of adoption on her neighbors' status \cite{granovetter1978threshold, watts2002simple, borge2013cascading, lopez2008diffusion, weng2013virality, bauch2013social, centola2007complex, eckles2018long, pinheiro2018consensus,hebert2015complex}. A spreading process with such a non-linear likelihood of adoption is a ``complex contagion'', whose properties can differ significantly from simple contagions \cite{guilbeault2018complex,centola2018behavior}. The spread of complex contagions is related intimately to the interplay of network structure and adoption patterns, relying on locally high prevalence and multiple peer influence in order to spread. 

\subsection{Relationship with past work}
The empirical evidence for complex contagions, including the propagation of online contagions, is accumulating \cite{weng2013virality,zhou2015impact,monsted2017evidence,romero2011differences,state2015diffusion,sprague2017evidence,steinert2017spontaneous} and several structural features influencing spread have been identified \cite{centola2015social,centola2018behavior,ugander2012structural,steinert2017spontaneous,guilbeault2021topological}. Beyond the adoption characteristics and network structure studied here, other factors influencing spread likely include individual heterogeneity, personal characteristics, strategic or reactive adoption, as well as global influences such as mass media \cite{state2015diffusion,bakshy2015exposure,toole2012modeling,traag2016complex}.

Threshold models \cite{krapivsky2011reinforcement} provide a simple and elegant way to capture non-linear adoption, which can be further generalized with dose response \cite{dodds2004universal,dodds2005generalized} and arbitrary adoption \cite{lopez2008diffusion} mechanisms. These models provide insights into how heterogeneous adoption thresholds \cite{granovetter1978threshold,watts2002simple} and the form of adoption functions interact with node degree on random networks. Assuming locally random tree-like networks (i.e. the absence of significant clustering), general conditions for global spread can be derived \cite{watts2002simple,dodds2011direct}. In some cases, the relevant micro-parameters of the model, such as the probability of adoption given one or two exposures, can be empirically measured to calibrate the model \cite{dodds2005generalized}. These models do not address the temporal dynamics of the contagion or connect its behavior to specific structural properties of the underlying network beyond the degree distributions. Moreover, these approaches do not study the dynamics and statistics of ``small'' contagions that never reach macroscopic size, and do not apply to community based or highly clustered networks. They do illustrate a subtle interaction between threshold level and degree heterogeneity that we build on in this paper: when an individual's adoption threshold is a function of the fraction (as opposed to the absolute number) of affected neighbors, low degree nodes are easily susceptible to be converted, but pass on the contagion to fewer neighbors. By contrast, high degree nodes are harder to activate but pass it on more widely. For a fixed average degree, it is therefore not immediately clear what the net effect of a wider degree distribution will be on the spread of such contagions.

The competing effects of clustering and ``long ties'' on complex contagions have been studied theoretically \cite{centola2007complex, montanari2010spread, nematzadeh2014optimal, eckles2018long} and empirically \cite{centola2010spread}. Game theoretic and threshold models have been used successfully to illustrate the key insight - supported by recent empirical work \cite{lehmann2018complex,hui2018scalable} - that clustering and communities can accelerate the spread of a complex contagion by allowing it to quickly reach locally high levels and spread one community at a time \cite{montanari2010spread,gleeson2008cascades}, whereas simple contagions converge faster for high-dimensional networks dominated by ``long ties'' \cite{eckles2018long}. Incidentally, similar insights emerge in the context of synergistic co-infections, whose coupled epidemiological dynamics also exhibit nonlinearities and thus complex contagion properties \cite{hebert2015complex}. These theoretical studies use approaches focused on deterministic mean field dynamics and convergence times, and are restricted to the regime of strong positive selection (i.e. where convergence is essentially guaranteed) \cite{montanari2010spread}.

\subsection{Overview of contributions}
{\J
The effects of general network features on the stochastic dynamics of complex contagions of a range of sizes (both the statistical distribution of rare events as well as the probabilities of global cascades) remain poorly characterized. Here we a present a framework based on mathematical tools and intuitions from population genetics to analyze these stochastic dynamics for arbitrary forms of complex contagions, and apply our model to understand the effects of key network properties including sparsity, community structure, and degree distributions. While the influence of these structural features has been illuminated previously \cite{centola2018behavior,guilbeault2018complex}, our approach builds on and supplements this prior work. 

Our method uses the language of population genetics to provide intuitive derivations of key properties of complex contagions and their dependence on the above network features. This approach allows us to analyze contagion dynamics at all scales of a network, from the local neighborhood to the community to the global scale, taking into account the interplay of ``selection'' (i.e. the local tendency for an idea to spread), diffusion (the random fluctuations in spread due to the stochastic nature of the process), and network structure. We study the contagions' full stochastic dynamics subject to arbitrary nonlinear adoption patterns and selection regimes, and we formulate network conditions under which complex contagions can reach global scales. 

A key idea is to use targeted approximations to derive an effective lower dimensional diffusion process that is (approximately) obeyed by the true contagion on the network. This approach highlights parallels between the competitions of genes in a population and the competition of memes in a world of minds and ideas. While our method is not necessarily applicable to arbitrary network structures, it provides insights in a variety of cases.}

\section{Our model}
In particular, we study here the fate and adoption of a newly arising idea on a network, giving rise to a complex contagion. We model this process in the framework of evolutionary game theory by considering individuals as the nodes of an undirected graph, with edges representing interaction and communication patterns (\fig{fig:real_networks_results} {\bf(a,b)}). We introduce the new idea as a single randomly chosen \ifc{} node on a network in which all other nodes are initially of \sus{}. Both types spread by contagion. In particular, we assume that individuals update their type as a continuous stochastic process, where the rate of switching depends on the fraction of neighbors of a given type: a \sus{} node becomes \ifc{} at rate 
$$r_1 = y \left[ 1 + f_1(y) \right]\;,$$
and \ifc{} nodes become \sus{} at rate 
$$r_2 = \left[ 1-y \right] \left[1 + f_2(y)\right]\;,$$
where $y$ is the local fraction of \ifc{} neighbors at a given node. For a complex contagion, $f_{1/2}$ are functions of $y$, while they are constants for simple contagions \cite{pastor2001epidemic, keeling1999effects, keeling2011insights, leventhal2015evolution}. Our main aim is to understand how successfully the new idea spreads through the network by calculating how the overall fraction of \ifc{} individuals, $\bar{y}(t)$, changes over time. {\J In a strict sense, we use $\bar{y}$ to refer to the overall (global) fraction of \ifc{} individuals and $y$ for the local fraction as seen by a given individual. When there is no possibility of ambiguity we will simply use $y$ in both cases for ease of notation.}

For concreteness we focus primarily on the simple illustrative case where $f_1(y) = \alpha y$ and $f_2(y) = \beta$, with positive $\alpha$ and $\beta$. This models ``positive frequency dependence'' \cite{pagel2019dominant}, where an idea is unpersuasive while rare but becomes more attractive as it is more widely adopted \cite{centola2007complex, montanari2010spread, nematzadeh2014optimal}. This is a natural assumption in many contexts (e.g. political views, preferences, games, or communication habits). However, we note that some ideas may be positively selected at all frequencies (i.e. negative $\beta$), in which case they will always tend to spread, and negative frequency dependence (i.e. negative $\alpha$) may also be relevant in other scenarios (e.g. fashion trends or baby naming). We further assume that $\alpha,\beta \ll 1$, which implies that the strength of selection is relatively weak, such that a preference for one or the other type only emerges on a collective population level (in the opposite case, the idea will tend to very quickly either spread or be eliminated). 

To some readers this model may appear reminiscent of SIS or SIR models in epidemiology \cite{allen2008introduction}, where the rate at which a susceptible individual becomes infected is often assumed to be proportional to the number of infected neighbors. Indeed, these models are encompassed by our framework. However, in SIS or SIR models the rate of recovery of an infected individual is generally not subject to neighbor influence, while the rate of spread is linear in the neighbors. This leads to simple contagion dynamics (with ``infected'' corresponding to \ifc{}) for low values of $\bar{y}$ and a diverging negative frequency dependent selection for large values of $\bar{y}$ (see the section ``Relation to epidemiological models'' in \cite{aps_cc_SI}). Therefore, small epidemics are well described with simple contagions, with the additional trivial consequence that large epidemics become exponentially unlikely. We do not study this case here. Instead, our paper is focused on the rich behavior resulting from positive frequency dependence once a sufficient prevalence $\bar{y}$ is reached. In this case, dynamics for low $\bar{y}$ are not well described with simple contagion models, considerations of social proof \cite{weng2013virality,monsted2017evidence} and evolutionary game theory are relevant, and the conclusions and intuitions gained from the model can differ substantially from those implied by epidemic models \cite{montanari2010spread}.

In \cref{fig:real_networks_results} {\bf (c-e)}, we explore how the spread of such a complex contagion is influenced by network structure. For this purpose, we consider the Facebook network from the Stanford Large Network Dataset collection \cite{snapnets}. We construct a sequence of networks with variable clustering but unchanged degree sequence by randomly swapping pairs of edges, and study contagions on this set of graphs. We find that the spread of simple contagions is largely insensitive to network structure (\cref{fig:real_networks_results} {\bf (c)}). By contrast, for complex contagions there is a critical level of clustering required to allow the contagion to spread globally. Below this level, the contagion becomes exponentially unlikely to fix across large networks. This can be seen in \cref{fig:real_networks_results} {\bf (c)} which shows that the fixation probability of the complex contagion is comparable to a simple contagion with negative selection when clustering is low but behaves like a simple contagion with positive selection as clustering gets sufficiently high. We also find that the contagion fixes one community at a time when clustering is sufficiently high (\cref{fig:real_networks_results} {\bf (d)}), but for moderate or low clustering values, all communities move through $y$ space more or less in unison (\cref{fig:real_networks_results} {\bf (e)}).

\subsection{Diffusion approximation}
{\J To quantify and analyze these effects, we begin by calculating the global rate at which \sus{} individuals become \ifc{}. In our model of contagion dynamics, this is 
\begin{widetext}
\begin{equation} \label{eq:expectation} \text{Rate}_{\text{A} \to \text{B}} = N (1 - \bar{y}) E_{A} [r_1(y)] = N (1 - \bar{y}) E_{A} [ y (1 + f_1(y))] = N(1-\bar{y})\left(E_{A}[y] + \alpha E_{A}[y^2]\right)\;. \end{equation}
\end{widetext}
{\J Here we use $E_{A}[\cdot]$ to denote the expectation value induced by the distribution of local $y$ as seen by a randomly chosen \sus{} individual, and equivalently for \ifc{}. The $N(1-\bar{y})$ term is the number of \sus{} individuals, and the expectation value gives the mean rate $r_1$ as averaged over all of these \sus{} nodes.  Through $E_A[r_1(y)]$, the rate crucially depends on the distribution of local $y$ seen by \sus{} individuals, which will depend on the network structure and the distribution of \ifc{} individuals on the network. The rate of the reverse process $\text{Rate}_{\text{B} \to \text{A}}$ has an equivalent form:
\begin{equation}\label{eq:expectation_b}
\text{Rate}_{\text{B} \to \text{A}} = N \bar{y} E_{B} [(1-y)(1+f_2(y)]\;.
\end{equation}
These transition rates define the stochastic process governing $\bar{y}(t)$, i.e. the total amount of \ifc{} individuals on the graph as a function of time. We will use the rates to develop an effective diffusion process describing its behavior. 

Let us consider $\delta \bar{y}$, the net change in $\bar{y}$ during some small time interval $\delta t$. The value of $\delta \bar{y}$ is determined by the difference between $A \to B$ and $B \to A$ transitions. The numbers of each of these transition events during a small time interval $\delta t$ can be viewed as independent poisson distributed random variables with rates as given by $\text{Rate}_{\text{B/A} \to \text{A/B}}$. Hence, the mean and variance of $\delta \bar{y}$ have the form
$$E[\delta \bar{y}] \equiv a(\bar{y}) \delta t =\frac{1}{N} \left( \text{Rate}_{\text{A} \to \text{B}} - \text{Rate}_{\text{B} \to \text{A}}\right) \delta t $$
$$Var[\delta \bar{y}] \equiv b(\bar{y}) \delta = \frac{1}{N^2} \left( \text{Rate}_{\text{A} \to \text{B}} + \text{Rate}_{\text{B} \to \text{A}}\right) \delta t\;.$$
For large $N$, we can treat $\bar{y}$ as a continuous variable between $0$ and $1$. The evolution of $\bar{y}$ can then be described by a Fokker-Planck equation \cite{ewens2012mathematical}
\begin{equation}\label{eq:diffusion}
\pd{f(\bar{y},t)}{t} = - \pd{}{y}\left(a(\bar{y}) f(\bar{y},t)\right)  + \frac{1}{2} \pdd{}{\bar{y}} \left( b(\bar{y}) f(\bar{y},t) \right)\;,
\end{equation}
where $a(\bar{y})$ captures selection and $b(\bar{y})$ captures diffusion strength. The process has absorbing boundary conditions at $\bar{y} = 0,1$ (since a population with all equal types will remain unchanged). We can summarize the behavior of this process with a selection pressure $s$, which we define in the standard way from population genetics \cite{ewens2012mathematical}, 
\begin{equation}\label{eq:s_def}
s \equiv \frac{2 a(\bar{y})}{N b(\bar{y})} = \frac{2 \left( \text{Rate}_{\text{A} \to \text{B}} - \text{Rate}_{\text{B} \to \text{A}}\right)}{\text{Rate}_{\text{A} \to \text{B}} + \text{Rate}_{\text{A} \to \text{B}}}
\end{equation}
This selection pressure determines whether the contagion will on average tend to grow $(s > 0)$ or shrink $(s < 0)$ and its magnitude measures the strength of selection as compared to the influence of random drift. 

The rates from equation \cref{eq:expectation} or equivalently the selection strength $s(\bar{y})$ from \cref{eq:s_def} define an effective diffusion process on the space of $\bar{y}$, as shown in \cref{eq:diffusion}. The properties of $\bar{y}(t)$ according to this process will mimic the properties of the true evolution of $\bar{y}(t)$ on the network.}

Thus, the key task for understanding the dynamics of the population is to find the local distribution of $y$ seen by individuals of different types, which allows us to compute the expectation values in \cref{eq:expectation} and hence the effective selection strength $s(\bar{y})$ from \cref{eq:s_def}. How the individuals are distributed among the network (and thus the local distribution of $y$) will depend on the network structure and the form of the functions $f_{1/2}(y)$. If the expectation values in \cref{eq:expectation} depend on additional degrees of freedom beyond the global value $\bar{y}$, then a higher-dimensional diffusion process (tracking more than just the global value $\bar{y}$ may be necessary to model the full dynamics on the graph accurately.}

\subsection{Selection regimes}
In a well-mixed population, where every node is connected to all other nodes, all individuals see the same global value of $y = \bar{y}$. Thus $E_A[y^2] = \bar{y}^2$, and hence 
\begin{equation}\label{eq:s_well_mixed}
s(\bar{y}) \approx \alpha \bar{y} - \beta
\end{equation}
in the limit where $\alpha,\beta \ll 1$. This simple linearly increasing form of $s(y)$ (omitting the bar for the rest of this section, since $y = \bar{y}$) is consistent with our model of an idea that is negatively selected when rare but that becomes more popular as it increases in frequency. The critical threshold frequency above which the idea becomes positively selected is $y = y_n \equiv \frac{\beta}{\alpha}$. In addition to this frequency dependence of $s$, the effect of random fluctuations is another key ingredient to understanding the behavior of the process. Standard results from population genetics \cite{ewens2012mathematical} imply that whenever the number of \ifc{} individual is small compared to the inverse of the selection pressure (i.e. when $N y |s| \ll 1$, in the illustrative case of constant $s$), the random stochasticity of the process dominates over the effects of selection, and the frequency of the idea is dominated by random ``genetic drift.'' By contrast, when $N y |s| \gg 1$, selection dominates over random drift, and the idea will tend to deterministically spread or be eliminated from the population.

We define $P_{reach}(y)$ as the probability that the contagion reaches a given value of at least $y$. This function captures the ability of the new idea to invade the population and describes the statistical behavior of the process at both small and large values of $y$. The selection regimes described above then define various different qualitative behaviors of $P_{reach}(y)$. When drift dominates, $P_{reach}(y)$ falls off as $\frac{1}{N y}$ as in a neutral random walk. In regimes of positive selection, a contagion reaching a given value of $y$ is almost certain to reach continuously higher values of $y$, so $P_{reach}$ is approximately constant. By contrast, when negative selection dominates, the contagion becomes exponentially less likely to reach ever higher values of $y$, so $P_{reach}$ falls off exponentially. 

In a complex contagion, where $s$ is a function of $y$, the process can encounter various such regimes of selection, as illustrated in figure \cref{fig:selection_regimes} {\bf (a-b)}. In our example where $s(y) = \alpha y - \beta$, the contagion begins with a neutral regime at low $y$. Depending on the total network size $N$, the contagion may then encounter a regime of negative selection before eventually reaching the regime of positive selection above frequency $y_n$ (with another regime of neutral selection in between where $s(y) \approx 0$). If the initial regime of negative selection is not too ``strong'', a contagion can ``tunnel through'' it by random chance, then encounter positive selection and fix.

{\J In the simple example of fixed selection, the boundaries between the regimes of selection are defined approximately by the points at which $N y |s| = 1$. In the more general frequency dependent case, we can use diffusion theory to generalize this condition (see ``Well mixed populations'' and ``Working with $NS(y)$'' in \cite{aps_cc_SI} for details). By placing a fictitious absorbing boundary at a given value of $y$, we can use the solution for the fixation probability of a diffusion process like \cref{eq:diffusion} with arbitrary $a(y)$ and $b(y)$ functions \cite{ewens2012mathematical} to derive
\begin{equation}\label{eq:p_reach_def}
P_{reach}(y)^{-1} \propto \int_0^y e^{-N S(z)} dz\;.
\end{equation}
with $S(y) \equiv \int_0^y \frac{2 a(z)}{N b(z)} dz = \int_0^y s(z) dz$. By inspecting \cref{eq:p_reach_def} and noting the exponential dependence, we can provide the generalized condition for transitioning between selection regimes:
\begin{equation}\label{eq:S_cond}
N |S(y) - S(y^*)| = 1\;,
\end{equation}
where $y^*$ is the argument of the most negative value of $S(x)$ reached for any value $x < y$. This elegantly generalizes the constant selection condition $N y |s| = 1$. The intuition behind the new condition is as follows. Consider the ratio 
$$\frac{P_{reach}(y)}{P_{reach}(y^*)} = \frac{\int_0^{y^*} e^{-N S(z)} dz}{\int_{0}^y e^{-N S(z)} dz}$$
which captures the scaling of $P_{reach}$ beyond the point $y^*$. How this quantity scales with $y$ depends how the value of $N S(y)$ compares to $N S(y^*)$. Because of the exponential, the largest value of the integrand dominates each integral. Thus, if $N S(y) \gg N S(y^*)$, the value of the integrand $e^{-NS(y)}$ in the denominator is negligible for $y > y^*$ and $P_{reach}(y)$ does not drop with $y$ and instead remains roughly constant in $y$ (positive selection). If $N S(y) < N S(y^*)$ (which implies $NS(y)$ is dropping with increasing $y$, otherwise there would be a different $y^*$), the integral in the denominator is dominated by the current value of $N S(y)$ and $P_{reach}(y)$ drops exponentially (negative selection). Finally, if $N S(y) \approx N S(y^*)$, the denominator grows roughly linearly with $y$ (neutral selection). Therefore, \cref{eq:S_cond} defines transition points between the various selection regimes, where $S(y) = \int_0^y s(z) dz$ captures the integrated effect of selection up to $y$. We illustrate the resulting selection regimes for our case of $s(y) = \alpha y - \beta$ in {\bf Supplementary Fig. 1} in \cite{aps_cc_SI}. Selection regimes are a key feature of a given contagion process as they allow an immediate high level description of its behavior.

\section{Random regular graphs}
\subsection{Approach}
To gain insight into the effect of various aspects of network structure on the spread of complex contagions, we now apply the ideas of effective diffusion processes and selection regimes to contagions on several archetypical families of networks.} One simple but critical aspect of network structure is that not all nodes are connected. To focus on the effects of this sparsity, we consider the spread of a contagion on a random regular graph, where each node is connected at random to exactly $k$ other nodes \cite{wormald1999models}. {\J In such a network, each node will no longer see the ``global'' value $\bar{y}$, but rather some local value that reflects the fraction of its neighbors that happen to be \ifc{}. In principle, determining these local values of $y$ is a complicated problem. However, because the network is random, we expect no strong locality in how \ifc{} individuals are distributed, so the neighbors of each individual form an approximately random sample of size $k$ of the whole population. This no-locality (or ``annealed'') \cite{derrida1986random, galstyan2007cascading} approximation is related to the assumption that a large randomly connected network initially looks ``locally tree-like'' \cite{watts2002simple,dodds2011direct} for a spreading contagion, but specifically ignores the fact that \ifc{} nodes are slightly more likely than chance to be connected to one another (this is because they can in reality only initially appear as a neighbor of another \ifc{} individual). The assumption of no locality contrasts with the case of a spatial network (e.g. a square lattice) where locality is fundamental to the network geometry (in this case the contagion becomes a front propagation problem and must be treated differently \cite{tanaka2017spatial}). We confirm the accuracy of the no-locality assumption in {\bf Supplementary Fig 2.} \cite{aps_cc_SI}, and contrast it with the case of spatial networks in {\bf Supplementary Figs. 3} and {\bf 4} \cite{aps_cc_SI}. 

In our approximation (see ``Sparse networks'' in \cite{aps_cc_SI} for details), the distribution of $y$ as seen by a given individual with $k$ neighbors follows a Hypergeometric (approximately a Binomial for $k \ll N$) distribution with success probability $\bar{y}$ and $k$ trials: 
$$y \sim \frac{1}{k} \text{Hypergeometric}(N,\bar{y}N,k)\;,$$
which implies $E_A[y] = \bar{y}$. In a simple contagion (with $f_{1/2}$ independent of $y$), only the first moment of the local distribution of $y$ appears in \cref{eq:expectation,eq:expectation_b}. A simple contagion is thus unaffected by network sparsity. By contrast, higher moments appear in \cref{eq:expectation,eq:expectation_b} for a complex contagion with $y$-dependent $f_{1/2}(y)$. Due to discreteness in the connectivity (and thus the nonzero variance in the distribution of local $y$), some \sus{} nodes will have more \ifc{} neighbors than others, and hence $E_A[y^2] > E_A[y]^2 = \bar{y}^2$. Sparsity therefore increases $\text{Rate}_{A \to B}$ and $s(y)$ compared to the well mixed behavior \cref{eq:s_well_mixed} and enhances the spread of a complex contagion.

\subsection{Results}
Using the hypergeometric distribution over local $y$ and its moments, we can obtain the expectation values in \cref{eq:expectation,eq:expectation_b} and hence compute the effective selection $s(\bar{y})$ on this graph using \cref{eq:s_def}. Specifically, we find that for large networks where $N \gg k$ (and assuming $\alpha, \beta \ll 1$), 

\begin{equation}\label{eq:s_eff_sparse} s(\bar{y}) = \alpha \left( \bar{y} + \frac{(1-\bar{y})}{k} \right) - \beta\;. \end{equation}

This reduces to the well-mixed solution $s(\bar{y}) = \alpha \bar{y} - \beta$ as $k$ becomes large, but for small $k$ selection is significantly enhanced, as shown in \cref{fig:selection_regimes} {\bf (c)}. The intuition is that for small $k$, some nodes will by chance happen to have a higher fraction of \ifc{} neighbors than others due to local sampling fluctuations. Because the transition rates increase non-linearly with $y$, the increased positive selection on the few individuals that see high values of $y$ outweighs the effect of the reduced value of $y$ seen by individuals with fewer \ifc{} neighbors. While this effect is present for all $k$, it becomes stronger for smaller $k$ since the variance in the locally observed $y$ increases with smaller $k$. 

The example of sparse regular networks illustrates several general patterns in our analysis. The distribution of \ifc{} individuals is influenced by the network structure and discreteness for any contagion process, but it is only for complex contagions that it affects selection and thus the spread.This happens through the higher moments of the distribution of local $y$, which only appear in \cref{eq:expectation,eq:expectation_b} if there is a frequency dependence of $f_{1/2}$, i.e. for a complex contagion. By contrast, as long as the first moment is unchanged from $\bar{y}$, a simple contagion is not affected by network structure (see ``Simple contagion'' in \cite{aps_cc_SI}).} 

Generally, for a given $\bar{y}$, structure influences how \ifc{} individuals are distributed during the contagion, which through \cref{eq:expectation,eq:expectation_b} interacts with the specific form of $f_{1/2}(y)$ to produce the effective selection strength $s(\bar{y})$. This determines regimes of selection and the overall behavior of the contagion. Moreover, $s(\bar{y})$ defines an effective diffusion process capturing the behavior of $\bar{y}(t)$, which we can easily solve using standard methods to obtain $P_{reach}(\bar{y})$, the fixation probability $P_{fix}$, properties of the temporal evolution \cite{eckles2018long}, or any other quantities of interest.  Thus we can reduce our problem to calculating the distribution of $y$ in the neighborhoods of \sus{} and \ifc{} individuals at a given global value of $\bar{y}$. In general, $s$ at any point in time will depend on the full configuration of the \ifc{} individuals on the network. However, using key assumptions about the dynamics, we can often significantly reduce the degrees of freedom on which $s$ depends. In the above example, by assuming no locality and noting the random connectivity of the network, we reduced the complexity of the process to a single degree of freedom: $\bar{y}$. 

\cref{fig:sim_results} {\bf b,c} shows that our theory accurately predicts the results of numerical simulations of the process for various degrees of sparsity. {\J Moreover, we show in \cref{fig:sim_results} {\bf b} that the simple condition $N |S(y) - S(y^*)| = 1$ accurately predicts transitions between selection regimes. In particular, the black arrows are the predictions for transitioning from initially neutral selection at small $y$ to negative selection, which is visible on the log-log plot as a change from a straight line to a downward bending shape of $P_{reach}(y)$. The white arrows are the predictions for transitioning from the negative selection regime to the positive selection regime (which manifests visually as a transition from a downward bending trend to flat $P_{reach}(y)$}.

{\J While a precise treatment of the additional effects of locality is beyond the scope of this work, we can provide some intuition for its effects. Locality slightly increases the chances of the extreme outcomes of having zero \ifc{} neighbors as well as the chances of having many \ifc{} neighbors (see {\bf Supplementary Fig. 2} \cite{aps_cc_SI}). This is because \ifc{} nodes are created by definition only if they are initially in contact with another \ifc{} individual, so they are slightly more likely than chance to be found next to each other. They are also more likely than chance to be connected to each other in a locally ``tree-like'' structure \cite{dodds2011direct}. Because the true distribution of $y$ is slightly wider than in our approximation, the variance is slightly higher and thus the effect on selection is slightly more positive than predicted. This explains the slight underestimation of $P_{reach}$ and $P_{fix}$ in \cref{fig:sim_results} by our approach. We have confirmed that these discrepancies disappear in a modified version of the simulation where node identities are shuffled on the graph at every time step (making the no locality assumption exactly true). As the specific form of the nonlinearity interacts with the distribution of $y$ through its higher moments, the differences in the distribution of $y$ compared to the no locality approximation could potentially lead to larger discrepancies between our theory and simulations for different nonlinearities. Nonetheless, the approximation allows us to build a quantitative and intuitive picture that captures important aspects of the true process.

\section{Community based networks}
\subsection{Approach}
Next we consider the effect of community structure, where the impact of within-community locality is essential to the contagion dynamics. } To analyze this effect, we consider random graphs that consist of randomly connected communities of $m$ individuals each. In particular, we assume every individual has exactly $k_i$ random connections within the community and $k_e$ outside of it, where $k_i + k_e \equiv k$. By tuning $k_i/k$, we can vary the strength of community structure. As $\frac{k_i}{k} \to 1$, we have very strong and cohesive communities, while $\frac{k_i}{k} \to \frac{m}{N}$ reduces to the case of a random regular graph of degree $k$.

{\J
We will provide a brief description of the approach, for more details we refer to ``Community based networks'' in \cite{aps_cc_SI}. To analyze the contagion on such a graph, we must understand how \ifc{} individuals distribute themselves across the network. For clarity, let use $z$ to denote the fraction of \ifc{} individuals within a given community. We are then interested in the distribution of the $z$ values, as seen across all communities in the network. Let us denote this distribution with $a(z)$, which gives the fraction of communities at a fixed value of $z$. Note that $z$ is discretized in units of $\frac{1}{m}$, and we have $\sum_z z a(z) = \bar{y}$. Because the connections on the network are random within and between communities, we will assume that each node sees a random sample of size $k_i$ from within the community with its internal edges, and a random sample of size $k_e$ of the rest of the graph with its $k_e$ external edges. This is effectively a targeted version of the no-locality assumption: for the same reasoning as with the regular random graph, while the distribution of node types across communities $a(z)$ matters, the location of \ifc{} individuals in a given community does not, and neither does how the communities are shuffled for a fixed $a(z)$. We demonstrate the validity of our assumptions in {\bf Supplementary Fig. 2} \cite{aps_cc_SI}. This allows us to determine the distribution of $y$ as seen by a given node:

\begin{equation}\label{eq:y_distribution_two_level}
y = \frac{i_i + i_e}{k_i + k_e}\;,
\end{equation}
where $i_i$ and $i_e$ are Hypergeometric random variables just like in the section on sparse networks representing the number of \ifc{}  neighbors coming from edges internal to the community and external to it, respectively. That is,
\begin{align}\label{eq:i_distribution_two_level}
i_i &\sim&\text{Hypergeometric}&(m-1,z m,k_i)\;, \nonumber \\ 
i_e &\sim&\text{Hypergeometric}&(N-m,N\bar{y} - z m,k_e)\;, 
\end{align}
Intuitively, in addition to discreteness effects as before, the distribution of $y$ for a given node is now a weighted mixture between the $z$ of the community that the node is located in, and the global value of $\bar{y}$. It is now more clear how the distribution $a(z)$ will affect the local distribution of $y$ as seen by a given individual: if the distribution $a(z)$ is tightly centered around the global $\bar{y}$, we expect the overall results to be very similar to a regular random graph of degree $k$, i.e. no significant effect of community structure. On the other hand, if the distribution $a(z)$ has significant departures from $\bar{y}$, (for example, most communities could be either ``full'' or ``empty'' and only spend little time in between), most nodes will either see very high values of $y$ or very low because of the partial effect of $z$ (which is modulated by the community strength $\frac{k_i}{k}$). This increases the variance in the distribution of $y$ (without affecting its mean), which similarly to the case of the regular random graph will change the effective selection on the graph through the higher moments appearing in \cref{eq:s_def,eq:expectation,eq:expectation_b}.

To find the distribution $a(z)$, we make the key assumption that for any given $\bar{y}$, the distribution of $y$ values seen within communities reaches a quasi-steady-state before $\bar{y}$ can change significantly across the whole graph. This distribution will depend on the connectivity of the network as well as the details of the transition probabilities. The steady-state approximation assumes that within-community dynamics are fast compared to global changes of $y$ across the whole network; we expect this to hold when selection is weak ($f_{1/2}(y) \ll 1$) and when communities are small and well-connected compared to the overall network. 

If we assume that we know the distribution $a(z)$, we can use the definitions of the contagion dynamics together with our knowledge of how the individual types are distributed to determine the rate at which $z$ changes in each community. Specifically, the rate of change of $a(z)$ for each value of $z$ will depend on the number of \ifc{} and \sus{} individuals in those communities ($mz$ and $m(1-z)$, respectively), as well as the rates at which individuals in communities of a given $z$ change types (which through \cref{eq:s_def} depend on their local distribution of $y$, which we can in turn obtain from \cref{eq:y_distribution_two_level,eq:i_distribution_two_level}). These transitions change the value of $z$ for a given community and thus cause transition rates between entries of $a(z)$ for neighboring values of $z$. This allows us to write down a nonlinear dynamical system for the temporal evolution of $a(z)$. By numerically finding the steady state of this system subject to the normalization conditions $\sum_z a(z) = 1$ and $\sum_z z a(z) = \bar{y}$, we can compute the equilibrium distribution for $a(z)$ (this ultimately becomes a nonlinear algebraic system of equations that can be solved using zero-finding routines, see ``Computing the equilibrium value of $a$'' in \cite{aps_cc_SI}).

The equilibrium distribution for $a(z)$ then allows us to compute the local distribution of $y$ as seen by a given node by using \cref{eq:y_distribution_two_level,eq:i_distribution_two_level} and the law of total expectation to marginalize over $z$ using $a(z)$. We show that our approximations accurately predict this distribution of local $y$ in {\bf Supplementary Fig. 5} \cite{aps_cc_SI}. As in the case of regular networks, the local distribution of $y$ implies an effective selection strength $s(\bar{y})$ acting on the contagion (\cref{fig:selection_regimes} {\bf (d)}). Overall, assuming that $a(z)$ is at equilibrium for any global $\bar{y}$ allows us to compute numerically an effective selection strength $s(\bar{y})$, which determines the behavior of the contagion. The agreement between our theoretical predictions and numerical simulations are shown in \cref{fig:sim_results} {\bf (d-f)}.

\subsection{Results}
When community strength is weak ($\frac{k_i}{k} \to \frac{m}{N}$), the equilibrium distribution of $a(z)$ is narrowly peaked around the global value of $\bar{y}$. In this case, each community simply behaves like a random sample of nodes from the overall network, and we have the same behavior as for the regular random network. By contrast, when communities are cohesive ($\frac{k_i}{k} \to 1$), the equilibrium distribution of $a(z)$ has the same mean, but is now more peaked at the extremes of $z=0$ and $z=1$. This ``U-shaped'' distribution of $z$ means that \ifc{} individuals are concentrated in just a few communities. The resulting distribution of local $y$ as seen by individuals is also more peaked at the extremes, since individuals see mostly edges from within their own communities, and those communities are either mostly \sus{} or mostly \ifc{}. This wider distribution of local $y$ enhances the spread of the contagion (for the same reason that higher variance in local $y$ enhances selection for the regular random graph). 

We provide here some intuition for the transition of $a(z)$ between the narrowly peaked and U-shaped regimes as a function of $\frac{k_i}{k}$. In \cite{aps_cc_SI} section ``Continuum approximation'' we provide a more quantitative justification based on an effective diffusion process for $z$ in a given community for fixed $\bar{y}$. For high $k_i$, the U-shaped distribution of $z$ arises because the many connections within a community can ``conduct'' influence between the types and thus cause rapid fluctuations of $z$ within the community, but only slow fluctuations between communities.} The rate of fluctuations are fastest when there are approximately equally many \ifc{} and \sus{} individuals in a community. By contrast, fluctuations are slow when nearly all the nodes within a community have the same type. The values of $z$ within a community (which are subject to random diffusion) will therefore spend most of their time at extreme values of $z \to 0$ or $z \to 1$. This intuition is confirmed in that we observe a critical level of community strength $\frac{k_i}{k}$ above which the equilibrium distribution of $z$ within a community turns from a narrow distribution (concentrated around the global $\bar{y}$ across the whole network) to a U-shaped distribution (same mean, but concentrated at the extreme values), as shown in {\bf Supplementary Fig. 5} \cite{aps_cc_SI}. The resulting variance in $y$ as seen by individuals is high, and selection is enhanced. Intuitively, it is much easier for the contagion to randomly reach a ``critical mass'' of popularity within a single community and experience positive selection there, compared to across the whole network. The contagion simply fixes one community at a time, as visualized in {\bf Supplementary Fig. 6} \cite{aps_cc_SI} as well as {\bf Supplementary Videos 1-3} \cite{aps_cc_SI}. These effects also explain our observations on the role of clustering and community strength on real social networks in \cref{fig:real_networks_results}. It is important to note that the unequal distribution of \ifc{} individuals among communities (just like the broader distribution of $y$ in sparse networks) is again a feature purely of the network structure and arises with or without complex contagion. However, it is only in the former case that this distribution has an effect on the spread.

\section{Graphs with variable degree distribution}
\subsection{Approach}
{\J Finally, we consider graphs with variable degree distributions and otherwise random connectivity. We present a brief description of the approach and refer to  ``Networks with degree distributions'' in \cite{aps_cc_SI} for details. Intuitively, there are competing effects and it is not immediately clear what the net impact of varying degree distributions should be on the spread of the contagion. On the one hand, high degree \ifc{} nodes are able to convert many other nodes once they are converted, but they are harder to convert themselves. On the other hand, it is easier to convert low degree nodes to \ifc{} for the same reason that low $k$ increases selection for the random regular graph, but those individuals in turn will influence fewer neighbors. Given a fixed average degree, it is not clear what effect a greater variance in degree will have.

In the case of non-regular graphs, the degrees $k$ of the nodes are distributed according to a degree distribution $P(k)$ (which for regular graphs has zero variance, an assumption that we now relax). For a given individual of degree $k$ on the graph, we will also need the distribution over the degrees $k'$ of their neighbors $P(k' | k)$. While this neighbor degree distribution can in principle be arbitrary, we expect it without further information to have the form $P(k' | k) \sim P(k') k'$ since each node of degree $k'$ has $k'$ edges to which one can be connected (any departure from this distribution is called ``assortativity''). 

For networks with a nontrivial distribution $P(k)$, it is no longer possible to calculate a selection strength $s$ that depends only on $y$. Instead, we must work with the fraction of nodes of each degree $k'$ that are \ifc{}, $y_{k'}$. This requires an explicit analysis of the fraction of \ifc{} individuals for each degree $k'$, which leads to a high-dimensional diffusion process. Note that this still reduces the effective degrees of freedom significantly compared to the true process on the network, but not as much as in the regular graph case where we only track a single degree of freedom. 

We can solve this multi-dimensional diffusion process using the no-locality approximation, i.e. assuming that nodes of degree $k$ see a random sample of all other nodes on the graph. The probability distribution of the value of $y$ seen by a given individual will now depend on the degrees $k'$ of the individual's neighbors through $y_{k'}$ (the probability of a given node being \ifc{} is $y_k'$ and depends on $k'$). The degrees of the neighbors $k'$ in turn depend on the neighbor degree distribution $P(k' | k)$.  Using the law of total expectation, we find the simple and intuitive result that nodes of degree $k$ see a distribution of $y$ identical to that for a $k$-regular random graph, with the global frequency $\bar{y}$ replaced by the ``effective frequency''
\begin{equation}\label{eq:z_k_def}
z_k = \sum_{k'} P(k' | k) y_{k'};.
\end{equation}
Using this distribution of the local value of $y$ as seen by a given node of degree $k$, we can use the same approach as for the regular graphs to determine the rates \cref{eq:expectation} and thus obtain the diffusion process. This time, however, there is such a process for each population of $N P(k)$ nodes at each value of $k$ and they are coupled together through the mixing across degrees in \cref{eq:z_k_def}. This coupled high-dimensional diffusion process in $y_k$ space must therefore be solved numerically.}

\subsection{Results}
To vary both the mean and variance of the degree distribution continuously, we consider graphs where the degree of each node is drawn from a $\text{Gamma}$ distribution with mean $k$ and variance $\sigma_k$. Specifically, in order to illustrate the effect of wide degree distributions, in \cref{fig:sim_results} {\bf (g-i)} we compare graphs with $\sigma_k = 0$ (i.e. regular random graphs as studied before) to networks with high degree variance ($\sigma_k = 30$) and equal mean degree. {\J We consider regimes that on a regular graph with would consist of initial positive selection ($s(0) > 0$), initial neutral selection ($s(0) = 0$) and initial negative selection followed by positive selection ($s(0) < 0$). Our theoretical predictions show excellent agreement with the full numerical simulations. Note that for graphs with high degree variance, the behavior of $P_{reach}(y)$ becomes ``less extreme'', whether selection is positive or negative (we find lower $P_{reach}$ in the case of positive selection and higher $P_{reach}$ in the case of initial negative selection). Overall, we find that broader degree distributions dampen the effects of selection (whether positive or negative) on the contagion, both for simple and for complex contagions.} Another effect is the consistent suppression of the contagion for very low $y$ (see \cref{fig:sim_results} {\bf (h)}), which is enhanced for distributions with significant degree correlations (see {\bf Supplementary Figure 8} \cite{aps_cc_SI}). We give intuition and a derivation for this effect in \cite{aps_cc_SI} (see the sections ``Suppression at low $y$'' and ``Impact of the neighbor degree distribution''). We also verify the soundness of our modeling approach by comparing the predicted local distribution of $y$ to observations in {\bf Supplementary Figure 7} \cite{aps_cc_SI} showing close agreement. 

\section{Phase transitions}
Whenever it is possible to compute an $s(y)$, our framework implies a simple condition under which the contagion can spread globally with finite probability even in arbitrarily large networks (i.e. global cascades are possible, see ``Phase transitions'' in \cite{aps_cc_SI}): the width of a region of negative $s(y)$ around $y=0$ must scale as $N^{-\gamma}$, with $\gamma \geq 1$. That is, the contagion must need to tunnel through at most a finite number of individuals to reach a frequency above which it is positively selected. Otherwise, the process encounters negative selection and is exponentially unlikely to spread globally for large $N$. {\J Using \cref{eq:s_eff_sparse} and setting $s(0) = 0$, this leads to the critical sparsity $k_{crit} = \frac{1}{y_n} = \frac{\alpha}{\beta}$ below which global contagion is possible (\cref{fig:pfix_phase_transition} {\bf(a)}). Note that this result is in line with previous work considering locally tree-like connectivity \cite{watts2002simple,dodds2011direct} and has a simple intuitive interpretation: each individual that sees at least one \ifc{} neighbor has $s(y) = s(\frac{1}{k}) \approx \frac{\alpha}{k} - \beta$. If this minimum selection is nonnegative, the contagion can spread globally.  

For community-based networks, we find that the effective selection strength has $s(0) = -\beta$, but jumps higher as $y \to \frac{m}{N}$ (see \cref{fig:selection_regimes} (d)). Global contagion is possible provided that $s(\frac{m}{N}) \geq 0$, because in that case the contagion only needs to overcome a fixed size negative selection regime of size at most $m$ that does not scale with $N$. Numerically, we find this implies a critical community strength $k_i/k$ above which complex contagions are able to spread globally by appearing popular and reaching critical mass in one community at a time, even though they do not have critical mass on the global network (\cref{fig:pfix_phase_transition} {\bf(b)}). This is in line with our initial simulations of contagions on real social networks \cref{fig:real_networks_results} {\bf (c-e)}.}

\section{Discussion}
{\J These results demonstrate quantitatively how interactions between non-linear adoption probabilities and network structure influence the dynamics and outcomes of complex contagions by modulating the effects of selection and stochasticity. A central idea was the use of targeted approximations (e.g. no locality on random networks, local vs. global equilibration time scales on community based networks) to reduce the contagion to an effective diffusion process on a lower dimensional space ($y$ for regular networks, $y_k$ for random graphs with degree distributions, and the space of per-community $z$ for community based networks) and hence obtain its statistical properties. This allows us to understand the behavior of both large and small contagions, as well as the emergence of global cascades.} These results help explain why the spread of even initially unpopular ideas and opinions can be enhanced both by overall sparsity as well as by cliques and other forms of community structure. They also show that in contrast to simple contagions (where the existence of highly-connected individuals always enhances spread), broad degree distributions dampen both positive and negative selection for complex contagions and hence have more subtle effects.







\begin{figure*}[!ht]
\noindent \makebox[\textwidth]{\includegraphics[width=1.0\textwidth]{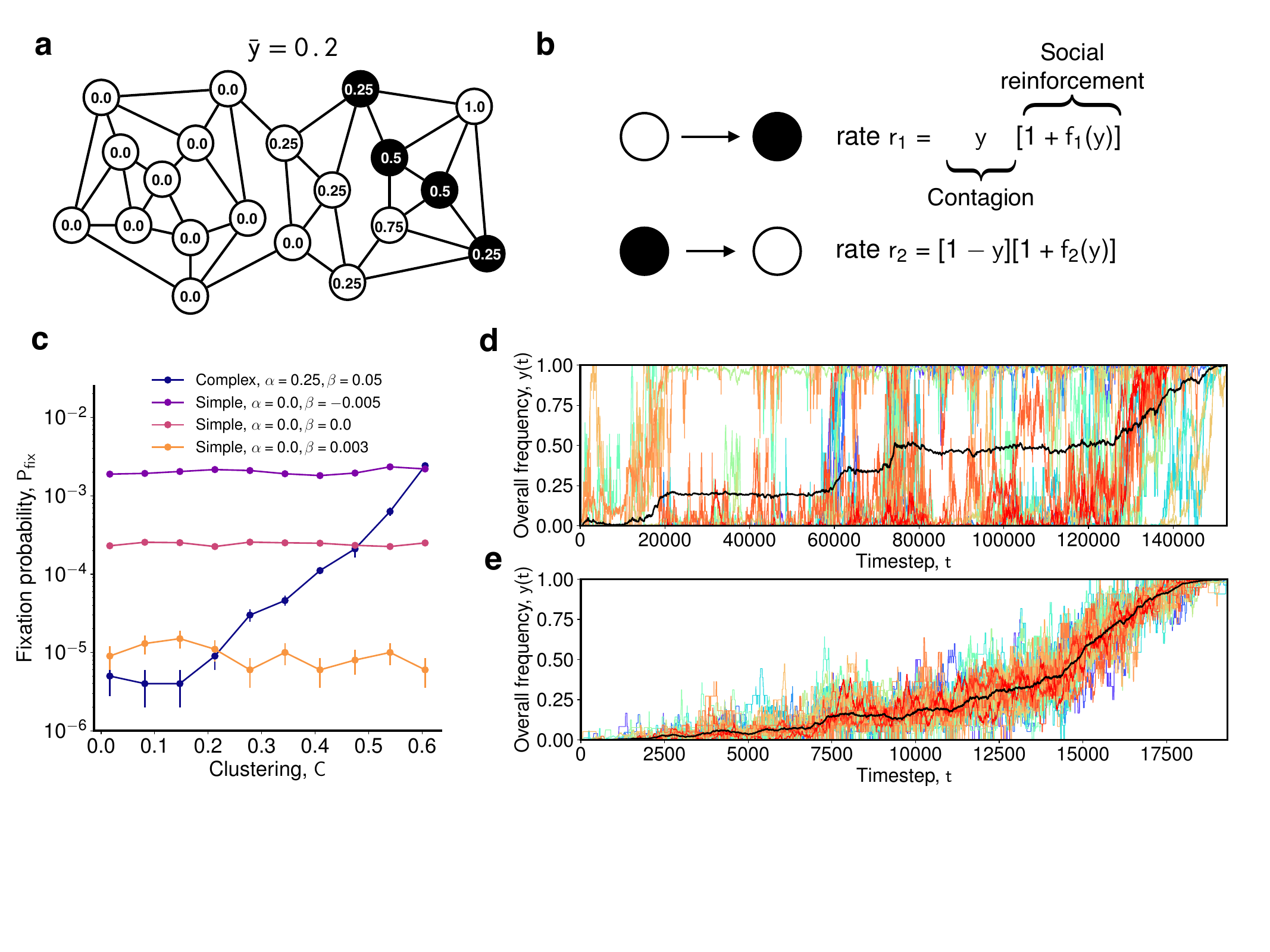}}
\caption{{\bf Model and simulations on real social networks.} {\bf a,} We model a complex contagion on a network where each individual can be \ifc{} or \sus{}. We denote the global frequency of \ifc{} individuals as $\bar{y}$. Each node sees a potentially different local fraction $y$ of \ifc{} neighbors (node labels). {\bf b,} Transition rates between \ifc{} and \sus{} individuals occur at rates $r_1$ and $r_2$; the form of $f_{1/2}(y)$ determines the non-linear adoption probabilities in complex contagions. {\bf c,} Simulations on networks of variable clustering derived by swapping pairs of edges in a Facebook network\cite{snapnets} ($N = 4039$, $k = 43$) show that the spread of complex (but not simple) contagions are highly sensitive to clustering. The line increasing with clustering is the complex contagion. The three flat lines correspond to simple contagions and are ordered top to bottom as in the legend. {\bf d,e,} Example frequency trajectories for contagions that fixed in our simulations. Each colored line shows the frequency within a given community as detected by a standard community detection algorithm \cite{raghavan2007near}, while the black line shows overall frequency $\bar{y}$. If the community structure is strong, the contagion fixes one community at a time, rapidly gaining and maintaining local popularity which helps the spread ({\bf d}, Clustering $C = 0.6$). If the community structure is weaker (but still detectable \cite{raghavan2007near}), the contagion instead spreads uniformly across the entire network ({\bf e}, $C = 0.2$). This is much less likely, so the fixation probability $P_{fix}$ is lower in this case.  Simulations assume $f_1(y) = \alpha y$, $f_2(y) = \beta$, $\alpha = 0.25$, and $\beta = 0.05$.}
\label{fig:real_networks_results}
\end{figure*}

\newpage

\begin{figure}[!ht]
\noindent \makebox[\linewidth]{\includegraphics[width=1.0\linewidth]{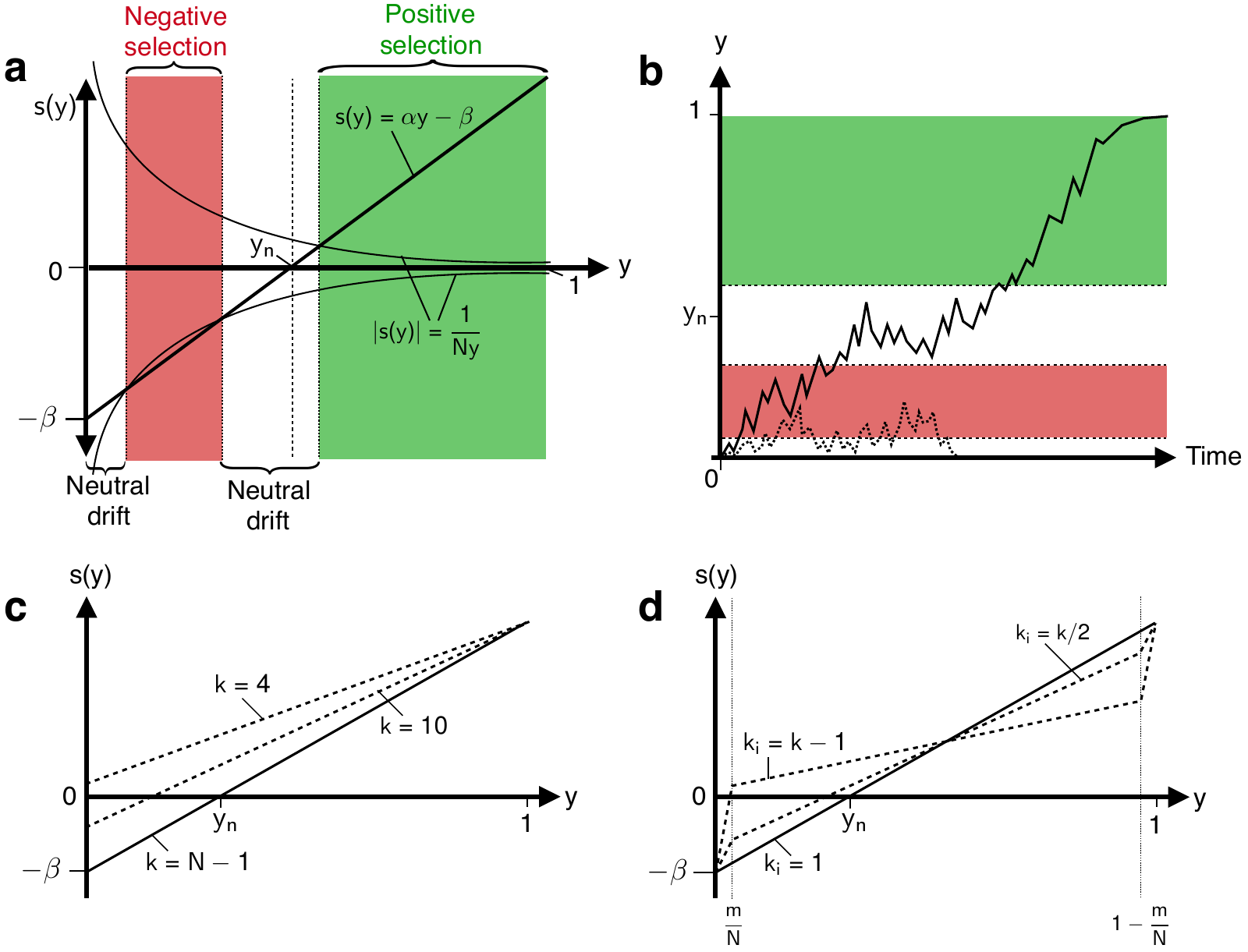}}
\caption{{\bf Selection and genetic drift in complex contagions.} For simplicity we omit the bar for $\bar{y}$ in all panels. {\bf a,} The condition $N |s(y)| y = 1$ distinguishes regimes where contagion dynamics are dominated by genetic drift, negative selection, or positive selection (this is an approximation to the exact condition $N |S(y)| = 1$, see \SI). {\bf b,} A contagion can spread globally if it reaches high enough frequency to be positively selected; this may require ``tunneling'' through a regime of negative selection at lower frequencies. {\bf c,d,} Sparsity ({\bf c}) and community structure ({\bf d}) can change the shape of $s(y)$ and hence alter the contagion dynamics. \label{fig:selection_regimes}}
\end{figure}

\newpage

\begin{figure*}[!ht]
\noindent \makebox[\textwidth]{\includegraphics[width=1.0\textwidth]{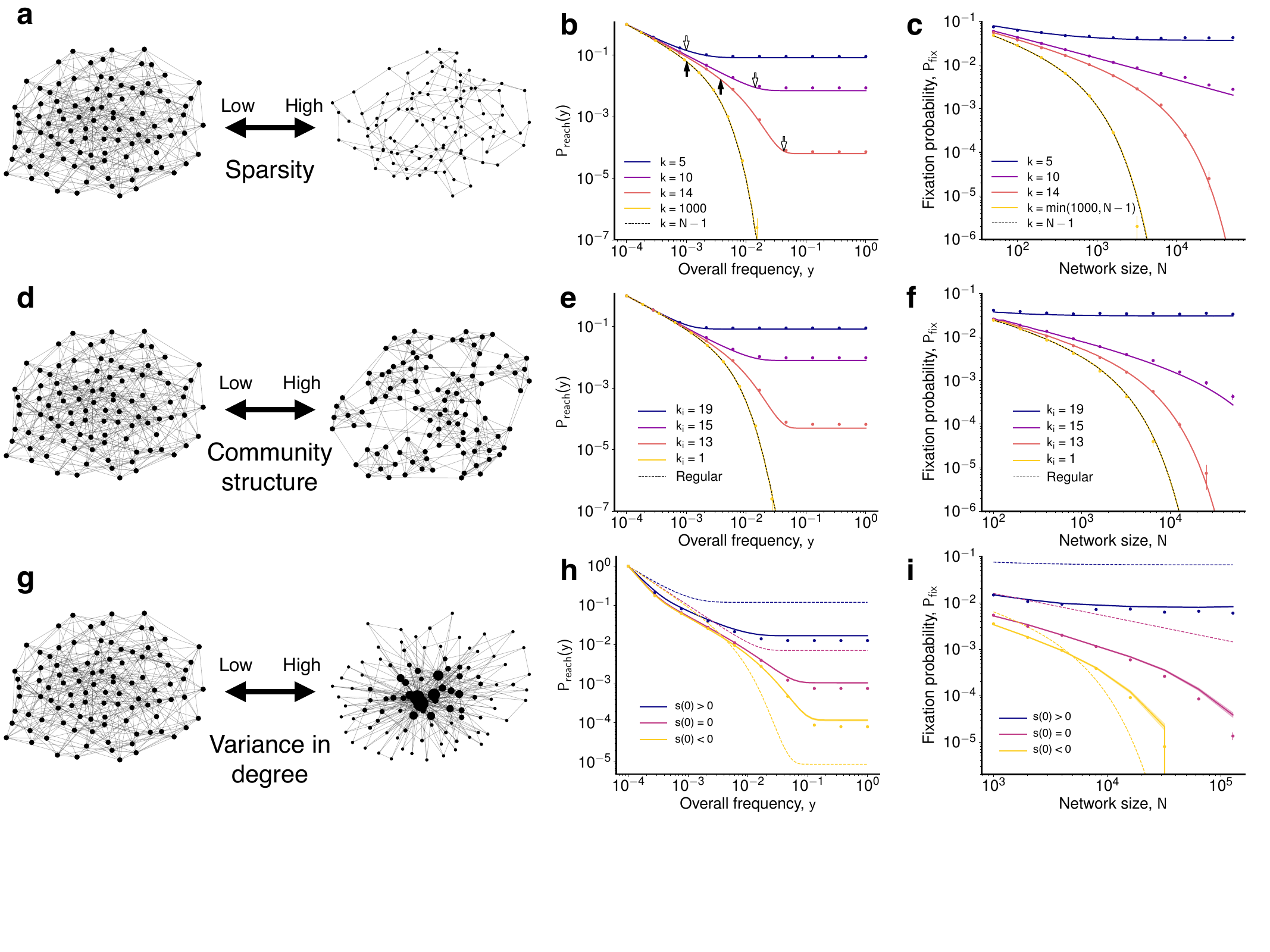}}
\caption{{\bf Network structure and the dynamics of complex contagions.} For simplicity we omit the bar for $\bar{y}$ in all panels.  {\bf a,} Illustration of networks that are more (right) or less (left) sparse. {\bf b,} Theoretical predictions (solid lines) and simulated results (for $N = 10,000$; dots) for $P_{reach}(y)$ for networks of different sparsity. Theoretical predictions for the transition to the regime of negative and positive selection are shown as black or white arrows respectively. {\bf c,} Theoretical predictions (solid lines) and simulated results (dots) for the fixation probability $P_{fix}(y)$ as a function of network size $N$. We show results for five values of $k$, corresponding to sparsity above (blue; top line), approximately on (purple; second line from top), and below (red; second line from bottom) the phase transitions allowing for global spread, as well as for a large value (yellow; bottom line) that approximates a fully-connected graph (dotted line). {\bf d,} Illustration of networks with more (right) or less (left) community structure. {\bf e,f,}  Theoretical predictions (solid lines) and simulated results (dots) for $P_{reach}(y)$ {\bf(e)} and $P_{fix}$ {\bf(f)} for networks with different strengths of community structure. {\bf g,} Illustration of networks with high (right) or low (left) variance in degree distribution. {\bf h,i,}  Theoretical predictions (solid lines) and simulated results (dots) for $P_{reach}(y)$ {\bf(h)} and $P_{fix}$ {\bf(i)} for networks with mean degree $k = 10$ and standard deviation in degree distribution $\sigma_k = 30$ for contagions with $\beta = 0.1$ {\bf(h)}, $\beta = 0.05$ {\bf(i)}, and three different values of $\alpha$ corresponding to initially positive, initially neutral, and initially negative selection on a regular graph with equal mean degree. Dotted lines show theoretical predictions for those equivalent regular graphs (i.e. with equal $k=10$ but $\sigma_k = 0$). Large $\sigma_k$ decreases $P_{reach}$ for positive selection and increases $P_{reach}$ for negative selection. In both cases, the absolute effect of selection is lessened by higher degree variance. Parameters: $\alpha = 1.0$, $\beta = 0.1$ {\bf(b)}, $\alpha = 0.4$, $\beta = 0.04$ {\bf(c)}, $\alpha = 0.88$, $\beta = 0.1$, $m = k = 20$ {\bf(e)}, $\alpha = 0.2$, $\beta = 0.025$, $m = k = 20$ {\bf(f)}, $\alpha = (2.5, 1.0, 0.71)$ {\bf(h)}, $\alpha = (1.25, 0.5, 0.36)$ {\bf(i)}. All parameters correspond to positive frequency dependence and are chosen so that the curves' distinguishing features are clearly visible within a reasonable range of magnitudes and computational budget. All line labels are ordered top to bottom in the legend in the same order as they appear in the plot itself. The dashed lines in {\bf(h-i)} have the same top to bottom ordering as the corresponding solid lines.\label{fig:sim_results}}
\end{figure*}

\newpage

\begin{figure*}[!ht]
\noindent \makebox[\textwidth]{\includegraphics[width=1.0\textwidth]{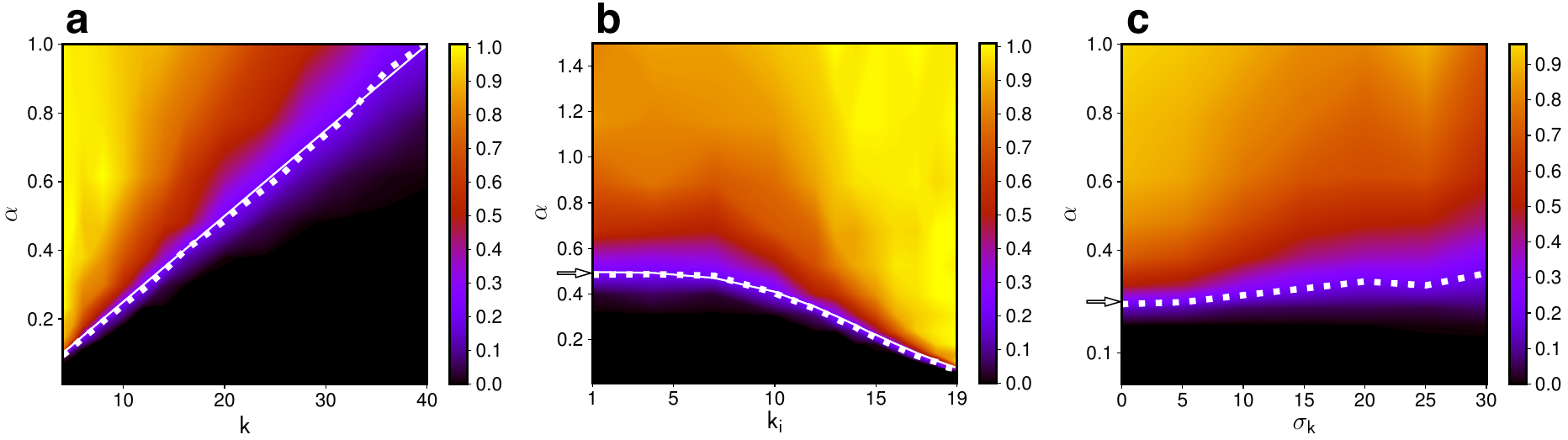}}
\caption{{\bf Phase transitions for complex contagions.} {\bf a,b,c,} Ratio of $P_{fix}$ on a network of size $N_1 = 50000$ to $P_{fix}$ on a network of size $N_2= 2000$ for contagions with $\beta = 0.025$, different values of $\alpha$ and varying sparsity {\bf(a)}, community structure ($k = m = 20$) {\bf(b)}, or degree distributions ($k = 10$) {\bf(c)}. Values close to one correspond to cases where $P_{fix}$ does not scale strongly with $N$, so global cascades are possible even in large networks. Solid white lines in {\bf(a,b)} denote the theoretically predicted phase transition, and the thick dashed white line indicates an observed ratio of $1/5 = \sqrt{N_1/N_2}$ (the empirical location of the phase transition). The theoretical value for {\bf(a)} is found by evaluating $s(0) = 0$ using equation \cref{eq:s_eff_sparse}. The theoretical value for {\bf(b)} is found by numerically evaluating $s(\frac{m}{N})$ and finding where it is equal to zero given the parameters (See the sections on ``Phase transitions'' in \cite{aps_cc_SI} for details). In {\bf(b)}, the location of the phase transition approaches the regular random graph value (white arrow, can be read off for $k = 20$ in panel {\bf(a)}) as the network loses community structure and becomes regular random ($k_i \to \frac{m}{N} \approx 0$). In {\bf(c)}, the empirical phase transition also correctly approaches the theoretical prediction (regular graph limit, white arrow) as $\sigma_k \to 0$. Since wider degree distributions weaken the effect of selection, the ``transition regime'' becomes noticeably wider for large $\sigma_k$.\label{fig:pfix_phase_transition}}
\end{figure*}
\clearpage

\clearpage
%

\end{document}
%



\title{Social network structure and the spread of complex contagions from a population genetics perspective: Supplemental Information}

\author{Julian Kates-Harbeck}
 \affiliation{Department of Physics,\\}
\author{Michael M. Desai}%
 \email{mdesai@oeb.harvard.edu}
\affiliation{%
Department of Organismic and Evolutionary Biology,\\ Harvard University, Cambridge MA~02138,~USA
}%

\date{\today}

\maketitle

\tableofcontents

\section{General analysis of complex contagions}

\subsection{Diffusion approximation}
We are interested in the evolution of $\bar{y} = \frac{n}{N}$ over time, where $n$ is the total number and $\bar{y}$ is the fraction of \ifc{} individuals in the population at a given time. During a short time step $\delta t$, we can use the rates
\begin{eqnarray}
r_1 &=& y \left[ 1 + f_1(y) \right] \nonumber \\
r_2 &=& \left[ 1-y \right] \left[1 + f_2(y)\right] \nonumber
\end{eqnarray}
to calculate the expected number of events on the network of \sus{} individuals switching to \ifc{}, $\delta n_+$. Since there are $N (1-\bar{y})$ total \sus{} individuals in the population, we have
\begin{equation}\label{eq:n_plus}
E[\delta n_+] = \delta t N \left[1 - \bar{y}\right] E_{A} \left[ y (1 + f_1(y))\right]
\end{equation}
where the expectation is taken over the distribution of $y$ as seen by all \sus{} individuals. Similarly, we have for the expected number of reverse event (\ifc{} individuals switching to \sus{})
\begin{equation}\label{eq:n_minus}
E[\delta n_-] = \delta t N \bar{y} E_{B} \left[ (1-y) (1 + f_2(y))\right] \;.
\end{equation}
The net change in $\bar{y}$ is $\delta \bar{y} = \frac{1}{N} \delta n \equiv \frac{1}{N} (\delta n_+ - \delta n_-)$ with expectation value $E[\delta \bar{y}] = \frac{1}{N}(E[\delta n_+] - E[\delta n_-])$. To lowest order in $\delta t$, since the $\delta n_{\pm}$ are independent binomial random variables, we also have
$$Var[\delta \bar{y}] = \frac{1}{N^2} (\delta n_+ + \delta n_-)\;.$$
For large $N$, we can treat $\bar{y}$ as a continuous variable between $0$ and $1$. As long as the expectations in \cref{eq:n_plus,eq:n_minus} depend only on the global value of $\bar{y}$ as the single degree of freedom, we can then follow the procedure in section 4.2 of Ewens\cite{ewens2012mathematical}, similar to the modeling of allele fates in population genetics, to find that the probability distribution $p(\bar{y},t)$ over $\bar{y}$ at a given time $t$ obeys the Fokker-Planck equation
\begin{equation}\label{eq:diffusion}
\pd{f(\bar{y},t)}{t} = - \pd{}{\bar{y}}\left(a(\bar{y}) f(\bar{y},t)\right)  + \frac{1}{2} \pdd{}{\bar{y}} \left( b(\bar{y}) f(\bar{y},t) \right)\;,
\end{equation}
with the definitions for the selection strength $a(\bar{y})$ and the diffusion strength $b(\bar{y})$:
$$E[\delta \bar{y}] \equiv a(\bar{y}) \delta t$$
$$Var[\delta \bar{y}] \equiv b(\bar{y}) \delta t \;.$$

Thus we can reduce the full dynamics on the graph to an effective diffusion process with the single degree of freedom $\bar{y}$ by computing the expectations $E[\delta n_+]$ and $E[\delta n_-]$ from \cref{eq:n_plus,eq:n_minus}. In particular, we have
\begin{equation}\label{eq:a_def}
a(y) = (1-\bar{y}) E_{A}[y (1 + f_1(y))] - \bar{y} E_{B} [(1-y) (1 + f_2(y))]
\end{equation}
and
\begin{equation}\label{eq:b_def}
b(y) = \frac{1}{N} \left[(1-\bar{y}) E_{A}[y (1 + f_1(y))] + \bar{y} E_{B}[(1-y) (1 + f_2(y))] \right]\;.
\end{equation}
Thus, the key task for understanding the dynamics of the population is to find the local distribution of $y$ seen by individuals of different types. This allows us to compute the above expectation values and find the effective diffusion dynamics governing the stochastic process. How the individuals are distributed among the network (and thus the local distribution of $y$) will depend on the network structure and the form of the functions $f_{1/2}(y)$. If the expectation values in \cref{eq:n_plus,eq:n_minus} depend on additional degrees of freedom beyond the global value of $\bar{y}$, then a higher-dimensional diffusion process may be necessary to model the full dynamics on the graph accurately (e.g. the section ``Networks with Degree Distributions'').

\subsection{Well mixed populations}
A well mixed population forms a special case of our model where every individual sees every other individual, i.e. the population is described by a complete graph. For large $N$, this means that every node sees the same, global $y = \bar{y}$ (we omit the bar for the rest of the section). In this case, the selection and diffusion strengths reduce to
\begin{equation}\label{eq:well_mixed_a}
a(y) = y (1-y) (f_1(y) - f_2(y))
\end{equation}
and 
\begin{equation}\label{eq:well_mixed_b}
N b(y) = y (1-y) (2 + f_1(y) + f_2(y))\;.
\end{equation}
Since our process is stochastic, we are interested in the statistical properties of the contagion. How large do these contagions tend to get? What is the probability for the contagion to reach a certain size? Let $P_{reach}(y)$ denote the probability that at least a fraction $y$ of the population becomes \ifc{} at some point during the contagion, starting with a single \ifc{} individual. This function captures the size statistics of the contagion. Since the only ``absorbing'' states of the process are $y = 0$ and $y = 1$, the contagion always dies out at $y = 0$ or fixes at $y = 1$ in the long run. We note that the temporal dynamics can also be obtained using this diffusion theory approach.

In section 4.3 of Ewens\cite{ewens2012mathematical}, for a diffusion process of the same form as in \cref{eq:diffusion}, results are derived for the fixation probability, i.e. the probability of reaching the absorbing state $y = 1$ for a given initial $y_0$. This calculation can be generalized to a fictitious absorbing boundary at an arbitrary value of $y$ (not necessarily equal to $1$). This allows us to calculate the probability of reaching that value of $y$ at least once, which is exactly $P_{reach}(y)$. For an initial fraction $y_0$ of \ifc{} individuals on the graph, we obtain
\begin{equation}\label{eq:p_reach}
P_{reach}(y ; y_0) = \frac{\int_0^{y_0} \psi(z) dz}{\int_0^{y} \psi(z) dz}\;,
\end{equation}
where
$$\psi(y) = exp\left(-2 \int_{0}^{y} \frac{a(z)}{b(z)} dz\right)\;.$$
This means that once we have the selection and diffusion strengths $a$ and $b$ for our effective process, we can directly calculate $P_{reach}(y)$. The functions $a$ and $b$ define the effective theory for any given network and selection functions $f_1$ and $f_2$.

While our results are valid for general frequency dependent functions $f_1$ and $f_2$, we will focus on the specific example of positive frequency dependence. Let us define the selection coefficient
\begin{equation}\label{eq:s_def}
s(y) = 2\frac{a(y)}{N b(y)}\;.
\end{equation}
This becomes 
\begin{equation}\label{eq:s_wm}
s(y) = 2 \frac{(f_1(y) - f_2(y))}{2 + f_1(y) + f_2(y)}
\end{equation}
for the well mixed population. In the case of weak selection where $s(y) \ll 1$ (also assuming that $|f_{1/2}(y)| \ll 1$), this reduces to $s(y) = f_1(y) - f_2(y)$. The function $s(y)$ captures to what degree selection favors \ifc{} over \sus{}. 

We will consider a simple form of positive frequency dependence, where $s$ is a linear function
$$s(y) = f_1(y) - f_2(y) = \alpha y - \beta\;.$$
In the limit of weak selection where $s(y) \ll 1$ and $|f(y)_{1/2}| \ll 1$, the exact forms of $f_1(y)$ and $f_2(y)$ are not important as long as their difference gives $s(y) = \alpha y - \beta$. We will chose the particular form $f_1(y) = \alpha y$, $f_2(y) = - \beta$. This corresponds to initial negative selection for small $y$. As more individuals adopt \ifc{}, positive selection takes over. This can be viewed as a classical situation for socially spreading phenomena, which often require a ``critical mass'' of adopters to become interesting or attractive. The critical threshold at which selection becomes positive is $y_n = \frac{\beta}{\alpha}$.

\subsubsection{Working with $N S(y)$}
Let us consider the effect of such frequency dependence on the success of contagions in a well mixed population. It is known\cite{ewens2012mathematical,desai2007beneficial} that in a population of size $N$, for a set of $n = y N$ individuals of constant fitness advantage $s$, there are two key regimes defined by the critical condition $n = \frac{1}{s}$. For $n \ll \frac{1}{s}$, random drift dominates the fate of the lineage (and $P_{reach} \sim \frac{1}{n}$). For $n \gg \frac{1}{s}$, the effect of the selection strength $s$ dominates (i.e. for positive $s$, $P_{reach}(y)$ is constant, and for negative $s$ it drops exponentially in $y$). 

The situation becomes more complex when $s(y)$ is frequency dependent. For instance, when $s(y) = \alpha y - \beta$, we must distinguish several different regimes (see {\bf Figure 2 (a-b)} in the main text). For small $y$, random drift dominates as in the case of constant selection. Then for larger $y$, negative selection can become significant. For still larger $y$, $s(y) \approx 0$ and random drift again dominates before $s(y)$ eventually crosses over into the positive selection regime at large $y$. If the initial negative selection regime is sufficiently ``strong'', most contagions will not make it to larger values of $y$. However, if that regime is not too strong, a contagion can ``tunnel through'' the negative selection regime, encounter positive selection, and fix.

To quantify this intuition, we can turn to \cref{eq:p_reach}. Using equation \cref{eq:s_def}, we define
\begin{equation}\label{eq:psi}
\psi(z) = e^{- N S(z)}\;,
\end{equation}
where
$$S(z) = \int_0^z s(y) dy\;.$$
For a fixed initial $y_0$ (one \ifc{} individual would correspond to $y_0 = \frac{1}{N}$), the numerator of \cref{eq:p_reach} is constant, and $P_{reach}(y)$ only depends on the denominator, such that $P_{reach}^{-1} \sim \int_0^y e^{- N S(z)} dz$. We are interested in the scaling of $P_{reach}$ with $y$. Because of the exponential, the value of the integral as a function of $y$ will be dominated by the maximum value of $- N S(y') \equiv \Sigma(y')$ over all values of $y' < y$. Let us denote with $M$ the highest local maximum of $\Sigma(y')$ for all $y' < y$, and with $y^*$ the value of $y'$ at which this occurs (which lies at $y' = \frac{1}{N} \approx 0$ unless there is a another higher local maximum). Note also that $P_{reach}$ is monotonically decreasing. We can then distinguish the following regimes of selection by their scaling behavior with respect to the point $y^*$:

\begin{itemize}

\item If $\Sigma(y) - M \gg  1$ (implying that $\Sigma(y)$ is increasing, otherwise there would be a larger local maximum) then the integral for any $y$ is always dominated by the current (maximal) value of $\Sigma(y)$. Thus $\frac{P_{reach}(y)}{P_{reach}(y^*)} \sim e^{- (\Sigma(y) - M)}$ drops exponentially in this regime, and we have negative selection.

\item If $|\Sigma(y) - M| \ll 1$ then for any $y$ the current value of the integrand has a comparable contribution to the overall integral as the last maximal value. This means the denominator is growing like $\sim (y-y^*)$ and $\frac{P_{reach}(y)}{P_{reach}(y^*)} \sim \frac{1}{y - y^*}$ which corresponds to neutral drift.

\item If $\Sigma(y) - M \ll -1$, then the current value of the integrand is not significantly contributing to the denominator, so $\frac{P_{reach}(y)}{P_{reach}(y^*)} \sim 1$ or $P_{reach} \sim \text{const}$. This corresponds to positive selection.
\end{itemize}

In other words, comparing $S(y) - S(y^*)$ to $\frac{1}{N}$ is the key and allows us to trace out qualitative regimes of selection starting from $y = 0$ by following the above distinctions. If $S(y) - S(y^*)$ is positive and large (compared to $\frac{1}{N}$), we have positive selection. If it is significantly negative (compared to $-\frac{1}{N}$), and decreasing, we have negative selection. If it is small in magnitude (compared to $\frac{1}{N}$), we have neutral drift. Whenever a new lowest local minimum of $S(y)$ is encountered, the definition of $y^*$ is reset. The process ``forgets'' about the behavior of $S(y)$ for $y < y^*$ as far as the further scaling of $P_{reach}(y)$ is concerned. Of course, other quantities such as the expected time for the process to reach a certain $y$ can still be affected. In summary, while for constant $s$ the critical comparison was between $|s| y$ and $\frac{1}{N}$ (selection scale vs. diffusion scale), the generalization to frequency dependent $s(y)$ becomes a comparison between $\left| \int_{y^*}^y s(z) dz \right|$ and $\frac{1}{N}$. We illustrate these regimes in {\bf Extended Data Figure 1}. In the case of well-mixed and sparse networks, where we have a closed form expression for $s(y)$, we can make a closed form prediction for the boundaries of the regimes, as shown with the arrows in {\bf Figure 3 b} in the main text. 

An interesting example of how this condition is new is the case of positive selection followed by negative selection. In this case $y^* = 0$ until the negative selection regime becomes wider than the inital positive selection. Despite the negative selection, the positive selection pushes the process forward so that it on average tunnels far into the negative selection regime, and only shows a drop in $P_{reach}$ once $NS(y) < -1$, i.e. when the total amount of negative selection outweights the forward push

In this work we consider the fate of a single new \ifc{} in a background of \sus{} individuals. However, this process can also be interpreted as the fate of an individual of a specific type in a background population of different competing types. Given positive frequency dependence, in a situation with a ``mutation term'', where new types are introduced at some constant rate, we would then observe a power law distribution of frequencies of adoption at low $y$, and a winner-takes-all behavior at high $y$, where one or a few types dominate a large portion of the population. This is precisely what is found in a study of music choices\cite{salganik2006experimental}, a situation where positive frequency dependence (making choices based on perceived popularity) may well play a role.

\subsubsection{Closed form fixation probabilities}
Understanding the boundaries of the various selection regimes allows us to make predictions about $P_{reach}(y)$. For the class of situations where $s(y)$ is monotonically increasing, we can distinguish the schematic situations shown in {\bf Extended Data Figure 1}. The parabolic shape shows $S(y)$, while the dashed lines show the values of $\pm \frac{1}{N}$. Recall that the condition $N S(y) = 1$ distinguishes the various selection regimes. For illustrative purposes, let us stick with the process where $s(y) = \alpha y - \beta$, such that $S(y) = \frac{\alpha}{2} y^2 - \beta y$. This defines a threshold $y_n = \frac{\beta}{\alpha}$ at which selection is neutral. If the initial negative selection is strong enough, the parabola can dip below the $\frac{-1}{N}$ line ({\bf Extended Data Figure 1 (a)}). The condition for this to happen is
\begin{equation}\label{eq:neg_selection}
 \frac{1}{2} \alpha N y_n^2 > 1\;.
\end{equation}
In that case, $P_{fix} \to 0$ exponentially with the value of $\alpha N y_n^2$. 

As the negative selection strength diminishes, the parabola may still go below $0$, but never actually dip below $\frac{-1}{N}$ ({\bf Extended Data Figure 1 (b)}). This means we have a neutral drift regime followed by positive selection starting at $y = y_p$, where the positive selection boundary $y_p$ is defined by $N(S(y_p) - S(y^*)) = 1$. Since neutral drift means that $P_{reach} \sim \frac{1}{N y}$, we know that $P_{reach}$ will drop like $\frac{1}{N y}$ until the value of $y$ where positive selection starts, and then remain constant. This means that $P_{fix} = \frac{1}{N y_p}$. In this case, we can distinguish three scenarios, depending on whether the initial selection $s(0)$ is negative ($y_n > 0$), exactly zero ($y_n = 0$), or positive ($y_n < 0$).

If we begin with negative $s(0) < 0$ (implying $y_n > 0$), the condition $S(y_p) - S(y_n) = \frac{1}{N}$ defines the point $y_p$ at which positive selection becomes dominant. Thus $P_{fix} = \frac{1}{N y_p}$. We obtain $y_p = \sqrt{\frac{2}{\alpha N}} - y_n$, and thus 
$$P_{fix} = \frac{1}{N y_n + \sqrt{\frac{2 N}{\alpha}}}\;.$$
If $s(0) \ge 0$, we know that $S(0) = 0$ and is increasing from there. Hence, $y_p$ is defined by $S(y_p) = \frac{1}{N}$. This gives
$$y_p = y_n + \sqrt{y_n^2 + \frac{2}{N \alpha}}\;,$$
and therefore
\begin{equation}\label{eq:pfix_pos}
P_{fix} = \frac{1}{N \left(y_n + \sqrt{y_n^2 + \frac{2}{\alpha N}}\right)}\;.
\end{equation}
Note that in this case, $y_n < 0.$

\subsubsection{Phase transitions}
Let us first consider the specific case where $s(y) = \alpha y - \beta$. For sufficiently large networks ($N \to \infty$), the condition for negative selection, \cref{eq:neg_selection}, will always occur if $y_n > 0$ (i.e. $s(0) < 0$). Thus, any such contagion will be exponentially unlikely to fix (though see the next paragraph for a more general discussion). By contrast, if $s(0) > 0$ (implying $y_n < 0$), we can Taylor expand \cref{eq:pfix_pos} for large $N$ to obtain
$$P_{fix} \approx \frac{1}{\alpha | y_n|}.$$
which crucially does not depend on $N$. In this regime, the contagion can fix with finite probability, regardless of the size of the network. For large $N$, the condition
\begin{equation}\label{eq:phase_boundary}
s(0) = 0
\end{equation}
thus defines the boundary between $P_{fix}$ dropping exponentially with $N$ ($s(0) < 0$), or remaining finite and constant ($s(0) > 0$). This is a phase transition between a regime where global cascades are impossible ($s(0) < 0$) and a regime where they are possible ($s(0) > 0$). Note that when $s(0) = 0$ exactly, we find that
$$P_{fix} = \sqrt{\frac{\alpha}{2 N}}\;.$$
Thus, a scaling of $P_{fix} \sim \frac{1}{\sqrt{N}}$ indicates that we are right on the phase boundary (as shown in {\bf Figure 4} in the main text).

Following this reasoning, we can obtain the following more general conditions. For arbitrary positive frequency dependence and $s(0) > 0$, global cascades are possible with constant probability. Moreover, if $s\left( y \ge \frac{n_0}{N}\right) > 0$ for some constant value of $n_0$ that does not scale with $N$ (even if $s\left(y < \frac{n_0}{N}\right) < 0$), then the impact of negative selection (if at all) will not scale with $N$, which means global cascades are still possible with finite and constant probability, regardless of the network size $N$. By contrast, if selection is negative over an initial regime that does scale with $N$, i.e. $s\left(y \le y_0 \right) < 0$ for some constant fraction $y_0$, global cascades are only possible if the population is small, with fixation probability scaling like $\sim N^{-1}$. For our specific example of $s(y) = \alpha y - \beta$, the critical population scale $N_{crit}$ for what it means to be ``small'' is $N_{crit} = \frac{2}{\alpha y_n^2}$ (from \cref{eq:neg_selection}). Otherwise, for sufficiently large $N$, negative selection will eventually cause fixation to become exponentially unlikely. Furthermore, if $s(0) = 0$ exactly, global cascades are possible, with fixation probability dropping at most like $N^{-1}$. To see why $P_{fix}$ can't drop faster under the assumptions of positive frequency dependence and $s(0) = 0$, note that in the limiting case of $s(y) = 0\;\forall\;y \ge 0$, we get neutral drift all the way which leads to $P_{fix} \sim N^{-1}$. Indeed, as shown above, for linearly increasing $s(y)$, the scaling is $\sim N^{-1/2}$. 

In summary, in the case of positive frequency dependence, global cascades are possible with finite probability for arbitrarily large $N$ if and only if there exists a constant value of $n_0$ (that does not scale with $N$) such that
\begin{equation}\label{eq:phase_boundary_general}
s\left( y \ge \frac{n_0}{N}\right) \ge 0\;.
\end{equation}

We later apply these insights to derive the phase boundaries for the sparse and community based networks (see {\bf Figure 4} in the main text), where we have closed form and numerical results for $s(y)$, respectively.

\section{Incorporating network structure}

\subsection{Simple contagion}
If $f_1(y)$ and $f_2(y)$ are independent of $y$ (i.e. for a simple contagion) we find that for regular graphs, both rates \cref{eq:n_plus} and \cref{eq:n_minus} are proportional (up to factors of $(1 +f_{1/2})$) to
$$ \delta t N(1-y) \frac{1}{N(1-y)} \sum_{i \in \sus{}} \frac{n_i}{k_i} = \delta t N y \frac{1}{N y} \sum_{i \in \ifc{}} \frac{\bar{n}_i}{k_i} = \delta t \frac{n_{AB}}{k}\;,$$
where $n_i$ is the number of \ifc{} neighbors of node $i$, $\bar{n}_i$ is the number of \sus{} neighbors of node $i$, $n_{AB}$ is the number of edges connecting \sus{} and \ifc{} individuals, and $k_i = k$ is the degree of the graph.
This means that 
$$a(y) = \frac{n_{AB}}{N k} (f_1 - f_2)$$
and
$$b(y) = \frac{n_{AB}}{N^2 k} (2 + f_1 + f_2)\;,$$
which implies constant selection
$$s(y) = \frac{2 a(y)}{N b(y)} = 2 \frac{f_1 - f_2}{2 + f_1 + f_2}\;,$$
which is equivalent to the well-mixed population (see \cref{eq:s_wm}).
This is the reason why network structure on regular graphs does not influence the contagion if it is simple. In a sense, due to the linear behavior of a simple contagion, the rates of change ``flow'' through all ``AB'' edges equivalently, regardless of where on the graph they lie. By contrast, for a complex contagion it matters how the AB edges are distributed.

It should be noted though that the temporal dynamics, which depend not only on the relative difference of the rates but also on their absolute value, can still be affected by network structure. Finally, for graphs with degree distributions, we find that the effects of both positive selection and negative selection are dampened (i.e. the behavior is moved closer to neutral) the wider the degree distribution is. Moreover, the contagion is universally suppressed at very small numbers of \ifc{} individuals (see ``Networks with degree distributions'' for an explanation of these small population size effects). 

\subsection{Sparse networks}\label{sec:sparse_networks}
Let us now consider the effect of population structure on complex contagions. We will consider the structures shown schematically in {\bf Figure 3} in the main text. To model sparsity, let us imagine a well mixed population (i.e. a random graph), where every node has only $k$ edges instead of $N-1$. We expect this model to reduce to the known well-mixed solution as $k \to N - 1$, with possible deviations as $k$ becomes small.

\subsubsection{No locality assumption}
To solve this model, we will assume that the random structure of the connections results in a situation where every individual simply sees a random sample of $k$ other members of the population as its neighbors. This is the so-called ``annealed approximation\cite{derrida1986random,galstyan2007cascading}''. Of course, since \ifc{} nodes can only emerge as neighbors of already \ifc{} nodes, there really are correlations in the localization of \ifc{} individuals near other such individuals. We will refer to such correlations as ``locality''. Locality will be most extreme on graphs with inherent local structure (such as a regular 1d or 2d lattice, see {\bf Extended Data Figures 3} and {\bf4}). By contrast, a random graph will have the least amount of locality ({\bf Extended Data Figure 2}).

In general, locality will result in values of $y$ as seen by individuals being slightly more extreme than the random graph assumption (since \ifc{} individuals are more likely to be next to each other). Since our assumption neglects the small amount of locality that exists even on the random graph, it underestimates the variance of the distribution of $y$ as seen by individuals, and thus underestimates $s(y)$ and $P_{reach}(y)$. We have verified that randomly shuffling \ifc{} individuals on the network at every time step (thus making the no locality assumption exactly true) removes the slight discrepancies between theory and simulation seen in {\bf Figure 3} in the main text. The assumption also overestimates the number of connections between individuals of different types. Since only such connections lead to changes in individual types and thus in $\bar{y}$, our approximation will underestimate the characteristic time scale on which $\bar{y}$ changes, i.e. the process will be slightly slower than our assumption predicts.

To get an intuitive picture of the effect of sparsity, consider a node with $k$ neighbors. If $k$ is large, due to the central limit theorem, the distribution of $y$ values as seen by nodes will approach a Gaussian distribution with variance decreasing for increasing $k$. The values of $y$ as seen by nodes with large $k$ will be tightly concentrated around the global value of $\bar{y}$. However, as $k$ becomes small, there is an increasing variance in the outcomes (see below paragraph for quantitative details), and thus a larger probability of a node simply ``by accident'' observing a high value of $y$ in its neighborhood. This node now experiences positive selection. Due to the nonlinearity in the selection function, this can influence the average selection on the graph as a whole. We quantify this intuition in the following section.

\subsubsection{Derivation of the effective selection $s(\bar{y})$}
According to our assumption, the number of \ifc{}  neighbors of a given \sus{} individual is equivalent to picking balls from an urn without replacement with a population size of $N-1$ and $\bar{y}N$ successes. Thus the value of $y$ as seen by this individual is distributed according to
$$y \sim \frac{1}{k} \text{Hypergeometric}(N-1,\bar{y}N,k)\;,$$
where the Hypergeometric distribution is parametrized by the population size, the number of successes, and the number of trials, respectively. If the individual is \ifc{}, the number of successes becomes $\bar{y}N -1$, since we know the individual itself is \ifc{} and there are no self-edges. Note that this distribution arises independently of the choices for $f_1$ and $f_2$ --- it is purely a function of the network structure. Its effect on the contagion will now depend on the form of $f_1$ and $f_2$. Specifically, for the choice of $f_1(y) = \alpha y$ and $f_2(y) = -\beta$, we can use \cref{eq:n_plus,eq:n_minus} to obtain
$$E_{A} [ y (1 + \alpha y)] = E_{A}[y] + \alpha E_{A}[y^2] = \bar{y} + \alpha (\bar{y}^2 + \bar{y} (1-\bar{y}) \frac{N - k}{N k})\;,$$
where we used the first and second moments of the hypergeometric distribution in the last equality and we have neglected terms of $O(\frac{1}{N})$. Similarly,
$$E_{B}[ (1-y) (1 - \beta)] = (1- \bar{y}) ( 1- \beta)\;.$$
Plugging into \cref{eq:a_def} and \cref{eq:b_def}, and using the definition of $s(y)$ (\cref{eq:s_def}), we obtain (to $O(s(y))$) the effective selection strength
\begin{equation}\label{eq:s_sparse}
s(\bar{y}) = \alpha \left(\bar{y} + \frac{(1-\bar{y}) (N-k)}{k N}\right) - \beta
\end{equation} 
where the second term in the parentheses is the departure from the well mixed case. Note that this term emerged because of the expectation value of $y^2$, which only entered due to the ``nonlinearity'' introduced by $f_1(y)$ being a function of $y$. Sparsity only actually affects selection in the context of frequency dependence. It is the interplay of network structure and complex contagion that allows this effect to emerge.

\subsubsection{Limiting cases and phase transition}
It is worth noting the limiting cases of \cref{eq:s_sparse}. As expected, the solution reduces to the expression for the well mixed case as $k \to N$. For large $N$ but $k \ll N$, we have $s(\bar{y}) = \alpha (\bar{y} + \frac{(1-\bar{y})}{k})) - \beta$. Depending on how the critical selection threshold $y_n = \frac{\beta}{\alpha}$ compares to $\frac{1}{k}$, the effect of sparsity may reduce the effect of negative selection at low $\bar{y}$, or even completely remove the negative selection regime (see {\bf Figure 2c} in the main text). Indeed, the phase transition condition \cref{eq:phase_boundary_general} in this case leads to $s(0) = 0$, which in turn implies the following critical sparsity condition for the possibility of global cascades on large networks:
$$k_{critical} = \frac{\alpha}{\beta} = \frac{1}{y_n}\;.$$
One way to interpret this condition is as follows. The effective selection for low $\bar{y}$ is simply
$$s(\bar{y}) = \alpha (\bar{y} + \frac{(1-\bar{y})}{k})) - \beta \to \frac{\alpha}{k} - \beta;.$$
The term $\frac{\alpha}{k}$ can be interpreted as an effective change in the local $y$ from $0$ to $\frac{1}{k}$. This is the case because $\frac{1}{k}$ is the minimum value of $y$ seen by any individual on the network neighboring at least one \ifc{} node. If the local effective selection seen by such individuals is not negative, global cascades are possible for arbitrary $N$. This relates our results to the conditions obtained by the more ``\ifc{}-centric'' approach using locally tree-like networks in previous work \cite{watts2002simple,dodds2011direct}.

\subsection{Community based networks}

A key aspect of real social networks is the presence of communities, i.e. groups of individuals that are more strongly connected within the community than they are to individuals outside of the community. These communities tend to have many internal connections and high clustering \cite{newman2012communities,yang2016comparative}. To understand the effect of community structure on social contagions, we will consider a simple and symmetric network model model that allows for tunable community strengths and sizes, but purposefully does not include any other features (such as degree distributions or differences in clustering throughout the network). We chose this model to analyze only the effect of community structure, removing the possibly confounding influence of other structural patterns.

\subsubsection{Description and motivation of the model}
In particular, consider a graph of $N$ individuals made up of equally sized communities of $m$ individuals each. Each individual has exactly $k$ neighbors, of which $k_i \le m$ are internal to the community, and $k_e$ are external to the community. When $\frac{k_i}{k} = \frac{m}{N}$, the graph is equivalent to a random regular graph with degree $k$ (since there is no significant excess of connections within communities compared to between communities). On the other hand, when $k_i$ becomes a large fraction of $k$, the communities become more tightly clustered and separated from the rest of the graph (see {\bf Figure 3} in the main text). The fraction $\frac{k_i}{k}$ is equivalent to the ``mixing parameter'' in past work \cite{yang2016comparative}. Note that every node on this graph (and every community) is statistically equivalent to all others. This symmetry makes this model easier to analyze.

\subsubsection{Equilibrium assumption} As with the sparse regular graph, the key to solving this model is to find the distribution of $y$ as seen by a given node. Consider the distribution $\v{a} = \{a_i\}$, $i \in \{1,...,m\}$, where $a_i$ is the number of communities with exactly $i$ \ifc{} individuals in them. Let $z \equiv \frac{i}{m}$ denote the fraction of \ifc{} individuals in the community. Knowing $a_i$ fully characterizes the distribution of \ifc{} individuals across the network. Our goal is thus to find the distribution $a_i$ for any given global $\bar{y}$ on the network. This allows us to compute the expectation values from \cref{eq:n_plus} and \cref{eq:n_minus}, which solves the system.

How will the \ifc{} nodes be distributed across the different communities? As the graph becomes more random and well mixed, i.e. as $\frac{k_i}{k}$ approaches $\frac{m}{N}$, we expect the communities to behave as random samples of $m$ nodes of the population. Thus, $a_i$ will have a Gaussian peak where $\frac{i}{m} = \bar{y}$. In the opposite extreme where $k_i \to k$, there are many more connections within a community than between communities. Since state changes are primarily mediated via connections between individuals of opposite types, we expect the population types within communities to change much faster than they spread between communities. In particular, this means that a community will go through rapid change as $i \sim m/2$, but will remain a long time in states where $i \sim 0$ or $i \sim m$. Thus, in this regime, we expect $a_i$ to be peaked at the extreme values of $i \to 0$ and $i \to m$ (see ``Continuum approximation'' for more details). In other words, the contagion either dies out or fixes quickly within a given community and then spends much time in those extreme states. This means that overall, the communities are mostly either empty or full of \ifc{} nodes.

To quantify this intuition, let us make the assumption that while individuals change type, the overall fraction $\bar{y}$ on the network remains constant. In other words, we are assuming that the values of $z = \frac{i}{m}$ within communities can change much faster than $\bar{y}$ can change globally on the overall network. The process will then tend towards an ``equilibrium distribution'' for $a_i$. We can find this equilibrium by enforcing a constant overall value of $\bar{y}$, and allowing individuals to change type until the distribution $a_i$ reaches a steady state. This steady state will then approximate the real distribution of $a_i$ for any given value of $\bar{y}$. We expect this approximation to work well when $m \ll N$ (such that the diffusion within a given community is much faster than $\bar{y}$ can change across the whole network) and $|s(y)| \ll 1$ (such that the time scale over which $\bar{y}$ changes significantly will be long compared to the time we spend ``fluctuating around'' a given $\bar{y}$, giving the distribution over $\v{a}$ time to equilibrate). We note that in the opposite case when selection is strong, the contagion will either become extinct or fix very rapidly anyways, so the precise equilibrium distribution will be less important for the contagion statistics.

\subsubsection{Equilibrium distribution derivation}
To compute the equilibrium distribution, we will consider the rates at which communities transition between different $i$ values. For instance, when an individual in a community of type $i$ (i.e. with $i$ \ifc{} nodes) transitions from \sus{} to \ifc{}  that community now becomes type $i + 1$. Since we know the number of \ifc{} nodes in any community and the total number of \ifc{} nodes in all other communities (and connections are random), we can compute the distribution of $y_i$ --- the values of $y$ as seen by nodes in a community of type $i$. This in turn provides the rates at which nodes change type. which allows us to write a differential equation describing the transitions of communities between types $i \in \{1,...,m\}$. Setting $\pd{}{t}(\cdot) = 0$ results in a nonlinear algebraic equation whose solution is the steady state of the dynamical system, which is the desired equilibrium distribution.

\subsubsection{Finding the distribution of $y_i$}\label{sec:y_distr_tl}
Given the equilibrium solution for $\v{a}$, we can compute the distribution of $y_i$. Consider a node in a community with $i$ total \ifc{}  individuals. This node has degree $k_e + k_i$. We know inside the community there are $i$ \ifc{}  individuals. Outside, there are $N\bar{y} - i$. Thus, we have 
\begin{equation}\label{eq:y_distribution_two_level}
y_i = \frac{i_i + i_e}{k_i + k_e}\;,
\end{equation}
where $i_i$ and $i_e$ are Hypergeometric random variables just like in the section on sparse networks representing the number of \ifc{}  neighbors internal to the community and external to it, respectively. That is,
\begin{eqnarray}\label{eq:i_distribution_two_level}
i_i &\sim& HG(m-1,i,k_i)\;, \nonumber \\
i_e &\sim& HG(N-m,N\bar{y} - i,k_e)\;,
\end{eqnarray}
where $i \to i-1$ for the distribution of $i_i$ if the node in question is \ifc{} (because we know the node itself is one of the \ifc{} individuals in the community). The equations above define the distribution of $y_i$ for any given node in a community of type $i$. We will denote expectations over this distribution with a subscript $i$. To obtain the overall expectation, we need to average over all the $i$. There are $(m-i)$ \sus{} nodes in a community with $i$ \ifc{}  individuals, and $a_i$ is the number of such communities. Thus, for any function $g$, we have
$$E_{A}[g(y)] = \frac{1}{N(1-\bar{y})}\sum_{i = 0}^m a_i (m - i) E_{A,i}[g(y)]\;,$$
and
$$E_{B}[g(y)] = \frac{1}{N \bar{y}}\sum_{i = 0}^m a_i i E_{B,i}[g(y)]\;,$$
which can be computed using $\v{a}$. This allows us to find the functions $a(\bar{y})$ and $b(\bar{y})$, and thus $s(\bar{y})$, for any given $\bar{y}$.

\subsubsection{Computing the equilibrium value of $\v{a}$} 
To find the equilibrium distribution of $\v{a}$, let us imagine a fictitious process that represents the real dynamics on the network, except that we hold the overall number of \ifc{}  individuals $N \bar{y}$ constant. To enforce this, we can artificially and uniformly increase all transition rates changing \sus{} to \ifc{} individuals by a constant factor $\gamma$ compared to the reverse process such that $\bar{y}$ remains constant.

Consider now a given configuration $\v{a}$ of \ifc{}  individuals. We can then compute the probability of a given node from a given community changing type. Specifically let $P_{+,i}$ be the probability that any given \sus{} node from a community with $i$ \ifc{}  individuals becomes \ifc{} in some time $dt$, and vice versa for $P_{-,i}$. This changes $a_i \to a_i - 1$, and $a_{i+1} \to a_{i+1} + 1$. Overall, we can write
\begin{equation}\label{eq:steady_state}
\d{a_i}{t} = \gamma P_{+,i-1} a_{i-1} + P_{-,i+1} a_{i+1} - \gamma P_{+,i} a_{i} - P_{-,i} a_{i}
\end{equation}
where we use the constant $\gamma$ to weight $A \to B$ transitions uniformly to enforce constant $\bar{y}$. The stationary solution can be found by setting the left hand side equal to zero. There is such an equation for every $i \in \{0,...,m\}$. We also must enforce that the that the total number of communities is
$$\sum_{i} a_i = m$$
and that the total number of \ifc{}  individuals is
$$\sum_{i} i a_i = N \bar{y}\;.$$
Finally, we need to exclude values of $\v{a}$ that are impossible due to the limited number of total \sus{} or \ifc{} individuals. All entries $a_i$ where $i > N \bar{y}$ or $(m-i) > N (1-\bar{y})$ must be identically zero. This is because in the former case, a community with $i > N \bar{y}$ would have more \ifc{} individuals than exist on the entire graph, and is thus impossible. Similarly, a community with $(m-i) > N (1-\bar{y})$ would have more \sus{} individuals than exist on the entire graph and is also impossible. Of course, since $i$ and $m-i$ are bounded $\in [0,m]$, this ``finite size effect'' is only is relevant for $\bar{y}< \frac{m}{N}$ or $(1-\bar{y}) < \frac{m}{N}$.

The coefficients $P_{+,i}$ and $P_{-,i}$ depend on $a_i$, $i$, $N$, $m$, $k_i$, $k_e$, and $\bar{y}$. The degrees of freedom are the $a_i$ and $\gamma$. Note that we have one more equation than we have variables. Overall, we have a set of nonlinear algebraic equations which can be solved for $\v{a}$.
The coefficients are
\begin{eqnarray*}
P_{+,i} &=& (m - i) E_{A,i}[y(1 + \alpha y)] \\
P_{-,i} &=& i (1 + \beta) E_{B,i}[(1 - y)]
\end{eqnarray*}
where $E_{A/B,i}$ denotes the expectation as seen by a node of type $A/B$ in a community of type $i$. The expectations over $y$ can be computed using \cref{eq:y_distribution_two_level} and (\ref{eq:i_distribution_two_level}) and are different for each $i$ (since they depend on how many \ifc{}  individuals are in the given community). The first term carries the number weighting. For instance, in the first equation, there are $a_i$ subgraphs in question, each of which has $m - i$ \sus{}  individuals, each of which has a rate of $E_{A,i}[y(1 + \alpha y)]$ of becoming \ifc{}.

\subsubsection{Qualitative behavior and phase transition}
Numerical solution of the above described system leads to solutions for $s(\bar{y})$ as shown qualitatively in {\bf Figure 2 (d)}. Compared to networks without community structure, $s(\bar{y})$ has the same limiting values and is still monotonically increasing. However, we find an initial and final regime of strong change in $s(\bar{y})$ for  $0 < \bar{y} < \frac{m}{N}$ as well as $0 < (1-\bar{y}) < \frac{m}{N}$, where the strength of those jumps grows with the strength of the community structure. The width of the regime is related to the size $m$ of the communities and thus does not scale with $N$. Therefore, if the jump in $s$ surpasses the negative selection regime, global cascades are possible. Using the phase transition condition \cref{eq:phase_boundary_general} with $n_0 = m$ (since $m$ does not scale with $N$), this leads to the following condition for the possibility of global cascades on large networks: 
$$s\left(\frac{m}{N}\right) = 0\;.$$

\subsubsection{Continuum approximation}
To get some intuition for the behavior of the equilibrium distribution of $\v{a}$ in various parameter regimes, it is useful to consider a continuum approximation of the process leading to diffusion in $\v{a}$-space. In particular, let us neglect the influence of $s$ (i.e. set $f_{1/2}(y) = 0$) and consider the limit of large $m$ (and continuous $i$). Consider then a single community with $i$ \ifc{} individuals. Assume the overall fraction of \ifc{} individuals on the graph is $y$. Let $z = \frac{i}{m}$ within this community. Moreover, let $k_i/k = \delta$ denote the level of community strength. We then obtain (using \cref{eq:y_distribution_two_level} and (\ref{eq:i_distribution_two_level}))
\begin{eqnarray*}
P_{+,i} &=& (1-z) E_{A,i}[y] \\
P_{-,i} &=& z E_{B ,i}[1 - y]\;,
\end{eqnarray*}
where $E_{A/B,i}[y] \approx \delta z + (1-\delta) y$. Assuming as above that $\bar{y}$ is nearly constant on the time scale of equilibrating the $\v{a}$ distribution and following the same procedure as leading up to \cref{eq:a_def,eq:b_def}, we obtain a diffusion equation in $\v{a}$-space with selection and diffusion strength functions
\begin{equation}
f_{a}(z) = (1 - \delta)(\bar{y} - z)
\end{equation}
and
\begin{equation}
g_{a}(z) = \frac{1}{m} \left[ 2 \delta z (1 - z) + (1- \delta)(\bar{y} (1-z) + z (1-\bar{y})) \right]\;,
\end{equation}
respectively. This gives rise \cite{ewens2012mathematical} to a steady state density
\begin{equation}\label{eq:steady_state_cont}
a(z) \propto \frac{1}{g_{a}(z)}e^{2 \int_0^z \frac{f_{a}(x)}{g_{a}(x)} dx}\;.
\end{equation}
where the constant of proportionality is chosen so ensure normalization over the range $z \in [0,1]$

Note that the selection strength $f_{a}(z)$ acts as a ``restoring force'' bringing the values of $z$ near those on the graph overall ($\bar{y}$). Its strength is highest when the community strength is lowest ($\delta = 0$). In that regime, the diffusion also biases the resulting distribution over $z$ towards the value of $y$ (second term in $g_{a}(z)$). Overall, when $\delta \to 0$, the distribution of $z$ will be concentrated (peaked) around $\bar{y}$.

By contrast, when community strength is high ($\delta \to 1$), the restoring force becomes smaller and the diffusion is strong everywhere except at the edges $z = 0,1$. The process thus spends much time at those extreme values, which causes the equilibrium distribution of $\v{a}$ to take on a U-shaped distribution that is peaked at the edges (see the $\frac{1}{g_{a}(z)}$ factor in the solution). Thus, depending on the community strength $\delta$, we expect either a distribution that is peaked at the edges, or one that is centered around the value of $\bar{y}$ on the overall graph. Balancing diffusion and selection suggests that this transition occurs near $\delta_{critical} \approx \frac{m - 1}{m}$ (see the following section for details).

This means that the U-shaped distribution just described mainly occurs at or near the highest possible community strengths $\frac{k_i}{k} \to \frac{m-1}{m}$. This is confirmed in our numerical simulations (see {\bf Figure 3} and {\bf Extended Data Figure 5}). This effect is similar to the transition of the allele frequency spectrum from concentrated to U-shaped as a function of decreasing mutation rate \cite{ewens2012mathematical}. Increasing $\delta$ reduces mixing of the population and behaves like a decreasing mutation rate. Note that these effects are purely a consequence of the community structure of the network and arise with or without frequency dependent selection. It is only in the presence of frequency dependent selection that it has significant consequences for the spread of the contagion (see main text). Note also that the condition derived here is only intended as an approximation and mostly presented for purposes of building intuition, and is not intended as an exact condition. Since $m$ is in reality small and finite, the transition occurs over a finite range and can affect the spread of the contagion significantly even if the threshold has not been crossed (see {\bf Figure 4}).

\subsubsection{Condition on the critical value of $\delta$}

Let us find the approximate value of $\delta$ that distinguishes the regimes of peaked vs. U-shaped steady state distributions over $z$. If we want a U-shaped distribution dominated by the effects of diffusion, then we require the ``selection'' term $e^{2 \int_0^z \frac{f_{a}(x)}{g_{a}(x)} dx}$ to not become too dominant compared to the diffusion term $\frac{1}{g_{a}(z)}$. Let us take the specific example of small $\bar{y} \to 0$ and consider the behavior of $a(z)$ (the qualitative argument proceeds similarly independent of this choice \footnote{Another way to proceed is to consider the neutral selection \cite{ewens2012mathematical} condition $m z s' = 1$, where $s' \equiv \frac{2 f}{m g}$ is the selection function in $z$ space, and then require the condition hold for values of $0 \lesssim z \lesssim 0.5$}). In the case of a U-shaped distribution, $a(z)$ should not drop too much as $z \to 1$.  We will assume that the diffusion terms $\frac{1}{g_{a}(z)}$ have similar order of magnitudes at $z \to 0$ and $z \to 1$ (note that $z= 0$ is an absorbing state so the smallest allowed value is $z = \frac{1}{m}$). Then we only need to check that the argument of the exponential does not become too negative.

We have
\begin{eqnarray*}
\int_0^1 \frac{f_a(z)}{g_a(z)} dz &=& m \int_0^1 \frac{(1-\delta) (-z)}{2 \delta z (1-z) + (1-\delta) z} dz\\
        &=& - m \int_0^1 \frac{1}{1 + \xi x} dx = -m \frac{log(1 + \xi)}{\xi}
\end{eqnarray*}
where $\xi = \frac{2 \delta}{1-\delta}$ and we performed a change of variables $x = (1-z)$. As expected, the integral is negative, which will cause a drop in the steady state probability distribution going from $a(0)$ to $a(1)$ due to the exponential. The only way the drop is not too large is if we require the magnitude of the integral to be $\lesssim 1$, which gives
$$\frac{m}{\xi} log(1 + \xi) \lesssim 1$$
Taking $log(1 + a) \sim O(1)$ (which will give a rather loose condition on how large $\xi$ and hence $\delta$ must be) leaves us with 
$$\frac{m}{a} < 1$$
or 
$$\delta > \frac{m}{m+2}\;.$$
Noting that $0 \le \delta \le 1$ by definition, we thus need a value of $\delta$ near its maximum value of $1$. Conversely, the fraction of external connections $1-\delta$ should be at most $O(\frac{1}{m})$. Plugging this value of $\delta \approx 1$ back into $g_a(z)$, we confirm that both $g_a(\frac{1}{m}) \sim \frac{1}{m}$ and $g_a(1) \sim \frac{1}{m}$ as we had assumed. Moreover, the logarithm will be $log(1 + \xi) \sim O(log(m))$, which means that even with these large values of $\delta$ we will expect a modest polynomial (but not exponential) drop from $a(0)$ to $a(1)$. 

Balancing the effects of diffusion and selection on the equilibrium distribution thus suggests that we need the fraction of internal connections $\delta$ to be near its maximum value of $\frac{m-1}{m}$ in order to observe a U-shaped distribution. 

Another way to develop a similar argument is to consider $\bar{y} = \frac{1}{2}$. In this case, we are interested in the ratio between $a(\frac{1}{2})$ and $a(0)$ at equilibrium. A ratio of $\sim 1$ will denote the transition regime between U-shaped and peaked distributions. We have
$$a(0) = \frac{1}{g_a(0)} = \frac{2 m}{1 - \delta}$$
and
\begin{eqnarray}
\frac{a(\frac{1}{2})}{a(0)} &=&  \frac{g_a(0)}{g_a(\frac{1}{2})} \text{exp}\left(m {\int_0^{\frac{1}{2}} \frac{(1-\delta)(\frac{1}{2} - z)}{2 \delta z(1-z) + (1 - \delta) \frac{1}{2}}}dz \right) \nonumber \\
 &=&\frac{2m (1-\delta)}{2m} \text{ exp}\left(2m {\int_0^{\frac{1}{2}} \frac{(\frac{1}{2} - z)}{4 \frac{\delta}{1-\delta)} z(1-z) + 1}}dz \right) \nonumber \\\
 &=& (1-\delta) \text{ exp}\left(2m \int_0^{\frac{1}{2}} \frac{x}{\frac{\delta}{1-\delta} (1-4x^2) + 1}dx \right) \nonumber \\
 &=& (1-\delta) \text{ exp}\left(\frac{m}{2} \int_0^{1} \frac{x}{\frac{\delta}{1-\delta} (1-x^2) + 1}dx \right) \nonumber \\
&=& (1-\delta) \text{ exp}\left(\frac{m (1-\delta)}{2} \int_0^{1} \frac{x}{1-\delta x^2}dx \right) \nonumber \\
&=& (1-\delta) \text{ exp}\left(\frac{m (1-\delta)}{4 \delta} \int_{1-\delta}^{1} \frac{du}{u} \right) \nonumber \\
&=& (1-\delta)^{1 -\frac{m (1-\delta)}{4 \delta}} \nonumber
\end{eqnarray}
where we used the substitution $x = (1 -2z)$ going to the third line and $u = 1 - \delta x^2$ going to the sixth line. Setting the ratio $\frac{a(\frac{1}{2})}{a(0)} \lesssim 1$ gives
$$\delta \gtrsim \frac{m}{m+4}$$
as our condition, which again confirms that $\delta$ should be near its maximum value in order to allow a U-shaped distribution for $a(z)$.

\subsection{Networks with degree distributions}
We will now study the effect of having a distribution of degrees on the graph, where the structure of the connections is otherwise random. In particular, let us assume degree distribution $P(k)$ (the probability that a given node has degree $k$) and neighbor degree distribution $P(k' | k)$ (the probability that a neighbor of a node with degree $k$ has degree $k'$).

\subsubsection{No locality assumption}
Similarly to the case of sparsity, we will assume ``no locality'', i.e. that every node of degree $k$ sees a random sample of size $k$ of the whole population. In a complex contagion, the probability of a node changing type can depend on its degree (see the section on ``Sparsity''). Moreover, the probability of seeing a certain local fraction $y$ will depend on the frequency of \ifc{} individuals among nodes of each degree $k'$, as well as the neighbor degree distribution $P(k'|k)$. For this reason, it will be necessary to keep track of the frequency of \ifc{} individuals for each degree $k$. This then allows us to find the expectations from \cref{eq:a_def,eq:b_def} and solve the model.

\subsubsection{Multi dimensional diffusion}
Let the \ifc{} frequency among nodes with degree $k$ be $y_k$. What is the distribution of $\tilde{y}_k$, the local frequency $y$ as seen by a node with degree $k$? To derive its distribution, we will make some simplifying assumptions which we will relax later. We first draw the degree of each neighbor independently from the appropriate neighbor degree distribution. Let $n^k_l$ denote the number of neighbors of degree $l$, given that the node in question has degree $k$. Then
$$n^k_l \sim \text{Multinomial}(P(l | k), k )\;,$$
i.e. $n^k_l$ follows a multinomial distribution with parameters $P(l | k)$ as the success probability, and $k$ trials. Recall that $P(l | k)$ is the neighbor degree distribution on the network. Given the degrees of the neighbors, we can use $y_l$ to draw the number of neighbors of a given degree which are \ifc{}. Let $i^k_l$ denote the number of \ifc{} neighbors with degree $l$, where the central node has degree $k$. It is distributed according to
$$i^k_l \sim \text{Binomial}(y_l,n^k_l)\;,$$
i.e. $i^k_l$ follows a binomial distribution with success probability $y_l$ and $n^k_l$ trials. Finally,
$$\tilde{y}_k = \frac{1}{k}\sum_{l} i^k_l\;.$$
The above expressions define the distribution of $\tilde{y}_k$. Using the law of total expectations, we arrive at the following simple expression for the first and second moments of the distribution:
$$E[\tilde{y}_k] = z_k$$
and
$$E[\tilde{y}_k^2] = z_k^2 + \frac{1}{k} z_k (1 - z_k)\;,$$
where $z_k = \sum_{k'} P(k' | k) y_{k'}$ is the expected value of $y_k$ as seen by a node of degree $k$ (i.e. the mean value of $y_k$ weighted by the neighbor distribution $P(k'|k)$). Thus, $\tilde{y}_k$ behaves as if it was distributed according to
\begin{equation}
\tilde{y}_k \sim \frac{1}{k} \text{Binomial}(z_k, k)\;.
\end{equation}
Note the interesting parallel to the result for the random regular graph, which is the same except that $\bar{y} \to z_k$ for each $k$ (the Hypergeometric becomes a Binomial for large $N$). This means that nodes with smaller $k$ again experience a larger variance in the local distribution of $y$ and thus larger selection.

The previous assumptions are not exact and we make the following minor modifications to these results for all our numerical tests:
\begin{itemize}
    \item The neighbors of a node are generated by drawing without replacement from a finite population of available edges of various degrees. Thus, the distribution of neighbor degrees is really a multivariate Hypergeometric distribution (where the success parameters are the number of edges on the graph connecting nodes with degrees $k$ and $l$ for each $l$. This number is in expectation given by $N P(k) k P(l | k)$). 
    \item Whether a given neighbor is \ifc{} or \sus{} is also drawn from all $N P(k)$ nodes without replacement, thus the distribution is really $i^k_l \sim \text{Hypergeometric}(N P(k) y_k, N P(k))$.
    \item The distributions $P(k)$ and $P(k' | k)$ are empirical distributions on the actual finite network, not the distributions from which the network itself was generated.
\end{itemize}
Note that these discrepancies vanish as $N \to \infty$. Finally, recall that we have assumed (just like on the random regular graph) that there is no locality in the process. Since a node can only become \ifc{} if one of its neighbors was \ifc{}, this is of course not exactly true. We have verified that the slight discrepancies between the simulation results and our predictions ({\bf Figures 3 h,i} and {\bf Extended Data Figure 8}, see especially for $y \to 1$) vanish when node positions are randomized (keeping their degree the same).

Overall, this multi-dimensional diffusion approach can be seen as a stochastic extension to prior work \cite{keeling1999effects}, where to account for the effect of degree distributions, the authors also had to keep track of the contagion in the multi-dimensional space spanned by each degree (albeit in a deterministic, ODE setting).

\subsubsection{Suppression at low $\bar{y}$}
Note that unlike all models considered so far, we find that degree distributions affect $P_{reach}$ even at very low values of $\bar{y}$ near $\frac{1}{N}$. We explain this effect here. Fundamentally, it is due to the fact that the first \ifc{} node in the contagion can have a range of degrees, and the probability of the contagion continuing is highly dependent on (nearly directly proportional to) the degree of the first node (it has an approximately equal chance of changing the type of every node connected to the first node, but the number of such neighbors is proportional to the degree of the first node).

Even for simple contagions where $f_1(y) = f_2(y) = 0$, we can observe this effect. To give intuition, consider the rate $r_{die}$ at which the first \ifc{} individual becomes \sus{} for a simple contagion (in this case the process dies out). This rate is $1$. Let us call the degree of the first \ifc{} individual $k_f$. The rate at which any neighbor of this first \ifc{} individual becomes \ifc{} is given by $y$, which is in this case $\frac{1}{k_n}$, where $k_n$ is the degree of said neighbor.  In expectation, this becomes $\sum_{k_n} P(k_n | k_f) \frac{1}{k_n}$. Assuming the neighbor degree distribution is related to the overall degree distribution as $P(k' | k) = P(k')\frac{k'}{\bar{k}}$ (with $\bar{k}$ the mean degree on the graph, which implies no degree correlations), we have that the rate of any neighbor becoming \ifc{} is $r_{live} = k_f \sum_{k_n} P(k_n)\frac{1}{\bar{k}} = \frac{k_f}{\bar{k}}$ where $k_f$ is the degree of the first \ifc{} individual. The rate is the sum of the rates that any of the $k_f$ neighbors becomes infected. The probability that the process reaches $2$ \ifc{} individuals is then the probability that any neighbor becomes converted to \ifc{} (rate $r_{live}$) before the first individual becomes \sus{} (rate $r_{die}$). Since all these events are independent and exponentially distributed, this means that 
$$ P_{reach}(\frac{2}{N}) = \sum_{k_f} P(k_f) \frac{r_{live}(k_f)}{r_{live}(k_f) + r_{die}} = \sum_{k_f} P(k_f) \frac{ \frac{k_f}{\bar{k}}}{1 + \frac{k_f}{\bar{k}} }\;,$$
where we are averaging the expression over the value of the initial degree $k_f$. This expression is equal to $\frac{1}{2}$ for any delta function degree distribution (which is why for small $s(y)$, $P_{reach}(\frac{2}{N}) = \frac{1}{2}$ for all network models without degree distributions). However, the concavity of the function $\frac{x}{C + x}$ implies that this expression always has a value of less than $\frac{1}{2}$ for any degree distribution $P(k_f)$ with nonzero variance. This is why for degree distributions with high variance, $P_{reach}(\bar{y})$ is noticeably suppressed at low $\bar{y}$. 

\subsubsection{Impact of the neighbor degree distribution}
As mentioned above, a simple approximation to the neighbor degree distribution is to simply to pick the node distribution and weight it for the fact that each node of degree $k$ has $k$ edges to which one can be connected. This gives as the ``expected'' neighbor distribution $P(k' | k) \sim k P(k')$
and thus
$$P(k' | k) = \frac{k' P(k')}{\bar{k}}\;,$$
where $\bar{k}$ is the mean degree of the network. Note that this expression is independent of $k$, which would make $z_k = z = \frac{1}{\bar{k}}\sum_{k'} P(k') k' y_{k'}$. For many real graphs, this approximation is very accurate. However, some real graphs have biases where nodes of a given degree are more likely than expected to be connected to high or low degree nodes. This (degree) assortativity (also called ``degree correlation'') is captured by the full neighbor degree distribution $P(k' | k)$.

In {\bf Extended Data Figure 8}, we show the results for $P_{reach}(\bar{y})$ on a network generated from a degree distribution with mean degree $k = 20$ and standard deviation $\sigma_k = 30$, restricted however to having only $2$ possible values for the degree, where the minimum degree is $3$. This creates a bimodal distribution (with possible degrees $k = 3$ and $k = 139$) with significant degree correlations when using the stub connect algorithm (see \Methods{}). We compare simulation results with predictions based on the actual neighbor degree distribution $P (k'| k)$, as well as with predictions based on the expected neighbor degree distribution (assuming no degree correlations). We find that taking into account the full neighbor degree distribution changes the results appreciably, and is necessary for approximating the simulations correctly. Note that as before, the remaining discrepancies for high $\bar{y}$ are due to the ``no locality'' assumption not holding exactly. This particular graph has a positive assortativity (i.e. high degree nodes connect more than expected to high degree nodes, and low degree nodes connect more to low degree nodes). This enhances the impact of the choice of the first node (see the previous section, but now high degree nodes are even ``better'' for the contagion), and thus causes an even steeper drop at very low $\bar{y}$ than the results for the expected neighbor distribution.

\section{Relation to epidemiological models}
Many epidemiological models in the literature are so-called SIS (or SIR) models\cite{keeling1999effects}. They assume a transition probability from the ``susceptible'' state (\sus{}) to the ``infected'' state (\ifc{}) proportional to the local fraction of \ifc{} neighbors, with proportionality constant $\lambda$. The reverse process happens with a constant recovery probability $\mu$. In our framework, for regular graphs, this corresponds to
\begin{eqnarray*}
r_1(y) &=& y \lambda \\
r_2(y) &=& \mu\;.
\end{eqnarray*}
Unlike our model, the ``susceptible'' state does not spread by contagion (this would require an additional factor as in $r_2(y) = \left[1-y\right] \mu$). While for small epidemics ($\bar{y} \ll 1$) both models are equivalent, the constant value of $r_2$ prohibits large epidemics from reaching $\bar{y} \to 1$. In particular, the absence of the factor of $(1-y)$ in $r_2(y)$ causes negative frequency dependence in the selection as $\bar{y} \to 1$:
$$s(\bar{y}) \sim f_1(\bar{y}) - f_2(\bar{y}) \sim \lambda - \frac{\mu}{(1-\bar{y})}\;.$$
While this model thus also has features of frequency dependent selection (and thus is not a ``simple contagion''), the frequency dependence here has qualitatively different consequences. Unlike the rich consequences of positive frequency dependent selection beyond a threshold value of $\bar{y}$ as studied in this paper, this negative frequency dependence simply inhibits the epidemic from reaching large values of $\bar{y} \to 1$.

In SIR models \cite{keeling1999effects}, there is an additional ``recovery'' state which is accessible with a constant rate after a node has become infected. A recovered node cannot become reinfected. If the \ifc{} individuals on the graph are concentrated together (as is the case for strong communities), spread of the epidemic is inhibited. This is because the global rate of nodes switching from \sus{} to \ifc{} is proportional to the total number of edges linking nodes of different types on the graph, whereas the global rate of recovery is proportional to the total number of \ifc{} nodes. In a sense, every edge between two \ifc{} nodes is ``wasted'' in that it doesn't contribute to the spread of the epidemic. Thus, strong community structure inhibits the spread of the epidemic. These insights supports past results  that higher clustering (i.e. stronger community structure) leads to smaller and inhibited epidemics in an SIR model that includes a recovery state\cite{keeling1999effects}.

We note that an edge based approach\cite{keeling1999effects} to analyze general graphs with a given clustering coefficient works for deterministic, simple epidemics because in that case each edge independently transmits the infection. We have verified that for our case of general complex contagions, such an approach works qualitatively (showing that increased clustering enhances the contagion), but fails quantitatively. This is because for complex contagions and the nonlinear dependence of adoption on the neighbor prevalence, edges are no longer independent. 

If nodes can have different degrees, many epidemiological models consider type change to be proportional to the number (not the fraction) of neighbors of the opposite type. For infections proportional to the absolute number, having higher degree nodes allows them to become infected with higher probability as well as spread the infection with higher probability. They become super-spreaders, and the overall dynamics of the epidemic are now dominated by the precise ``long-tail'' statistics of high-degree nodes \cite{lopez2008diffusion,pastor2001epidemic}. Epidemics with transmission proportional to absolute number always spread better with higher mean degree and higher degree variance.

\section{Methods}

\subsection{Numerical Methods}

\subsubsection{Analytical predictions of contagion statistics}
For the sparse network we can obtain an analytical solution for the function $s(\bar{y})$, while this function is the solution of a numerical procedure for the community based networks (see ``Community based networks'' in the \SI{}). In both cases, we then numerically evaluate the appropriate standard integrals from diffusion theory \cite{ewens2012mathematical} to obtain solutions for $P_{reach}(\bar{y})$.

For the degree distribution network, we explicitly run a simulation of the multi-dimensional diffusion process (see the section ``Multi dimensional diffusion'' in the \SI{}). For each simulation run, we sample the degree sequence from the full degree distribution. We also choose the initial degree of the first \ifc{} individual at random from the degree distribution. From then on, for every time step, we calculate $y_k$ --- the frequency of \ifc{} individuals among nodes of degree $k$ --- for all $k$. Given $y_k$, we calculate the rates of switching \sus{} and \ifc{} individuals of a given degree. We sample the number of switching events from a binomial distribution, where the number of individuals constitutes the number of trials, and the switching rate multiplied by a small time step (chosen such that the success probability is $< 0.1$) constitutes the probability of success. We have verified that the results of the simulation do not change noticeably with a smaller time step.

\subsubsection{Network simulations}
In order to simulate the process on real networks, we generate random graphs according to the structural features in question. We then perform a large number of simulations to obtain statistics on $P_{reach}(\bar{y})$ and $P_{fix}$. In particular, each simulation begins with a network of all \sus{} individuals, with a single randomly placed \ifc{} individual. We then update the types of all nodes according to a Gillespie algorithm, where the rates are given by the rates $r_1$ and $r_2$ from the main text. The algorithm is terminated when the absorbing states of $\bar{y} = 0$ (extinction) or $\bar{y} = 1$ (fixation) are reached.

\subsubsection{Local distributions of $y$}
In {Extended Data Figures 2-5}, we compare the predicted distributions of $y$ as seen by individual nodes to those observed in simulations, for a given global value of $\bar{y}$. Since the global value of $\bar{y}$ varies over time in simulations, we run simulations as usual starting with $\bar{y} = \frac{1}{N}$, and use data from all nodes during all time steps where the global value of $\bar{y}$ is equal to the desired value. Data is collected for $20$ separate simulation runs.

\subsection{Generating random networks}

\subsubsection{Real social network with variable clustering}
For {\bf Figure 1 (c-e)}, we construct a sequence of networks based on the Facebook network ($N = 4039$, $k = 43$) from the Stanford Large Network Dataset collection \cite{snapnets}. By considering random $A-B, C-D \to A-C, B-D$ swaps, we reduce clustering until a desired value is reached, while keeping the degree sequence intact. To test the impact of this reduction in clustering on the community structure of the network, we measure ``community overlap'' between the original and modified networks, by perform a bipartite matching of communities found by a standard community detection algorithm \cite{raghavan2007near}. We find that the communities still overlap to $80\%$ when clustering is reduced from the original value of $0.6$ to $0.2$. Community overlap finally drops to below $0.2$ as clustering is reduced to that of a random network. 

\subsubsection{Community based network} For the community based network model, random graphs are generated such that in the resulting graph every node and community in the network is statistically equivalent. Every community has exactly $m$ individuals, and every individual has exactly $k_i$ random connections to nodes within its community, and $k_e$ random connections to nodes outside the community, where $k_i + k_e \equiv k$. In other words, the network structure is a regular random graph of communities, where each community is itself a regular random graph. First each community is sampled as a regular random graph. Then the connections between communities are sampled on a community basis by sampling the supergraph of community connectivity as a regular random graph (with duplicate edges allowed) with $m k_e$ edges per community. Then the edges incoming to each community are evenly distributed among its member individuals. The resulting inter-community connectivity pattern is then rewired randomly without changing the degree sequence (by considering random $A-B, C-D \to A-C, B-D$ swaps) until all edges are valid (i.e. no duplicate edges).

\subsubsection{Variable degree distributions} Random graphs are generated by sampling a degree sequence from the specified degree distribution. In our simulations, we choose a $\text{Gamma}$ distribution with a given mean $k$ and variance $\sigma_k^2$. These degrees are then matched up with the stub connect algorithm \cite{arman2021fast}. The resulting connectivity pattern is then rewired randomly without changing the degree sequence (by considering random $A-B, C-D \to A-C, B-D$ swaps) until all edges are valid (i.e. no duplicate edges).

\subsubsection{Lattice networks}
The lattice networks in {\bf Extended Data Figures 3} and {\bf 4} are generated as linear (1D) and square (2D) lattices with periodic boundary conditions. On the 1D lattice, we connect each node to its $k$ nearest neighbors. On the 2D lattice, we connect each node to its closest neighbors ($k = 4$), or all its second neighbors ($k = 24$).

\section{Future work}
A promising direction for future work that goes beyond the scope of this paper would be an attempt to analyze the effects of multiple structural features interacting (such as degree distributions and community structure), which could allow the quantitative application of the framework to ``real'' networks, perhaps by fitting their macroscopic structural properties. This raises interesting questions about what properties would be most important to capture, and how they might be combined in a diffusion based analysis. While this would certainly be a worthwhile extension, we believe that our study of the various effects in isolation provides important initial insights into the dynamics and interactions at play.

\section{Code and Data availability}
All simulations and numerical calculations were performed with Julia 1.1. Our code is open source and available at \url{www.github.com/jnkh/epidemics}. The network data used is publicly available \cite{snapnets}.

\renewcommand{\thetable}{\arabic{table}}   
\renewcommand{\thefigure}{\arabic{figure}}
\renewcommand{\figurename}{Supplementary Figure}
\renewcommand{\tablename}{Extended Data Table}
\setcounter{figure}{0}
\setcounter{table}{0}
\crefname{figure}{\textbf{Supplementary Figure}}{\textbf{Supplementary Figures}} 
\Crefname{figure}{\textbf{Supplementary Figure}}{\textbf{Supplementary Figures}} 

\section{Supplementary Figures and Videos}

\paragraph*{\bf Supplementary Video 1} The progression of a contagion that fixes on a community-based network with low community strength. The contagion fixes uniformly across all communities. The visualization is as in {\bf Figure 1 d-e}. Transparent nodes are \sus{}; solid nodes are \ifc{}. Parameters: $N = 100$, $\bar{k} = 20$, $k_i = 4$.
\paragraph*{}

\paragraph*{\bf Supplementary Video 2} The progression of a contagion that fixes on a community network with high community strength. The contagion fixes one community at a time. Transparent nodes are \sus{}; solid nodes are \ifc{}. Parameters: $N = 100$, $\bar{k} = 20$, $k_i = 19$.
\paragraph*{}

\paragraph*{\bf Supplementary Video 3} The progression of a contagion that fixes on a real social network with strong clustering and cohesive communities \cite{snapnets}. The contagion fixes one community at a time. Transparent nodes are \sus{}; solid nodes are \ifc{}. Node size scales with degree. Parameters: $N = 357$, $\bar{k} = 10$.

\begin{figure*}[!ht]
\noindent\makebox[\linewidth]{
  \includegraphics[width=0.99\linewidth]{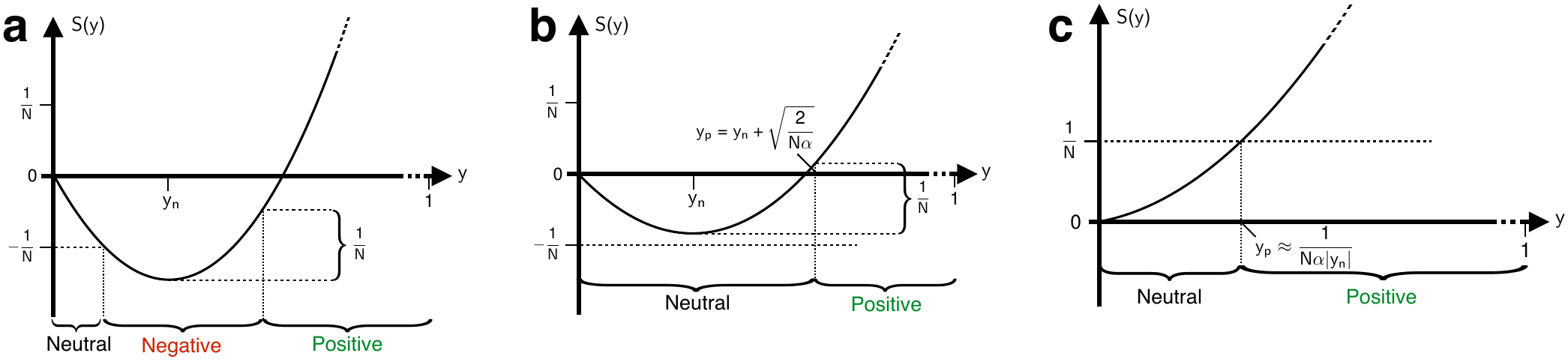}
 }
\caption{{\bf Scaling regimes for positively frequency dependent complex contagions.} For simplicity we omit the bar for $\bar{y}$ in all panels. The quantity $N S(y)$ determines the various selection regimes. A linearly increasing selection strength of the form $s(y) = \alpha y - \beta$ leads to a quadratic $S(y) = \frac{\alpha}{2} y^2 - \beta y$ which we can then compare to the relevant scale $\frac{1}{N}$. There are three possibilities: {\bf a,} $S(y)$ dips below the $-\frac{1}{N}$ line and thus enters a regime of negative selection. Once the difference from the lowest point of $S(y)$ to the current position becomes $\frac{1}{N}$, the selection becomes positive (see ``Working with $N S(y)$'' for details). {\bf b,c,} $S(y)$ never dips below the $-\frac{1}{N}$ line, so the process experiences neutral drift until $S(y)$ grows by $\frac{1}{N}$ from its minimum value. This happens at $y = y_p$, at which point positive selection takes over. It follows that $P_{fix} = \frac{1}{N y_p}$. If $S(y)$ dips below $0$ initially ({\bf b}), the fixation probability scales like $N^{-1}$. Otherwise ({\bf c}), the fixation probability does not scale with $N$, and global cascades are possible for arbitrarily large networks.}
  \label{edfig:edfig1}
\end{figure*}

\clearpage

\begin{figure}[!ht]
 \noindent\makebox[\linewidth]{
  \includegraphics[width=1.0\linewidth]{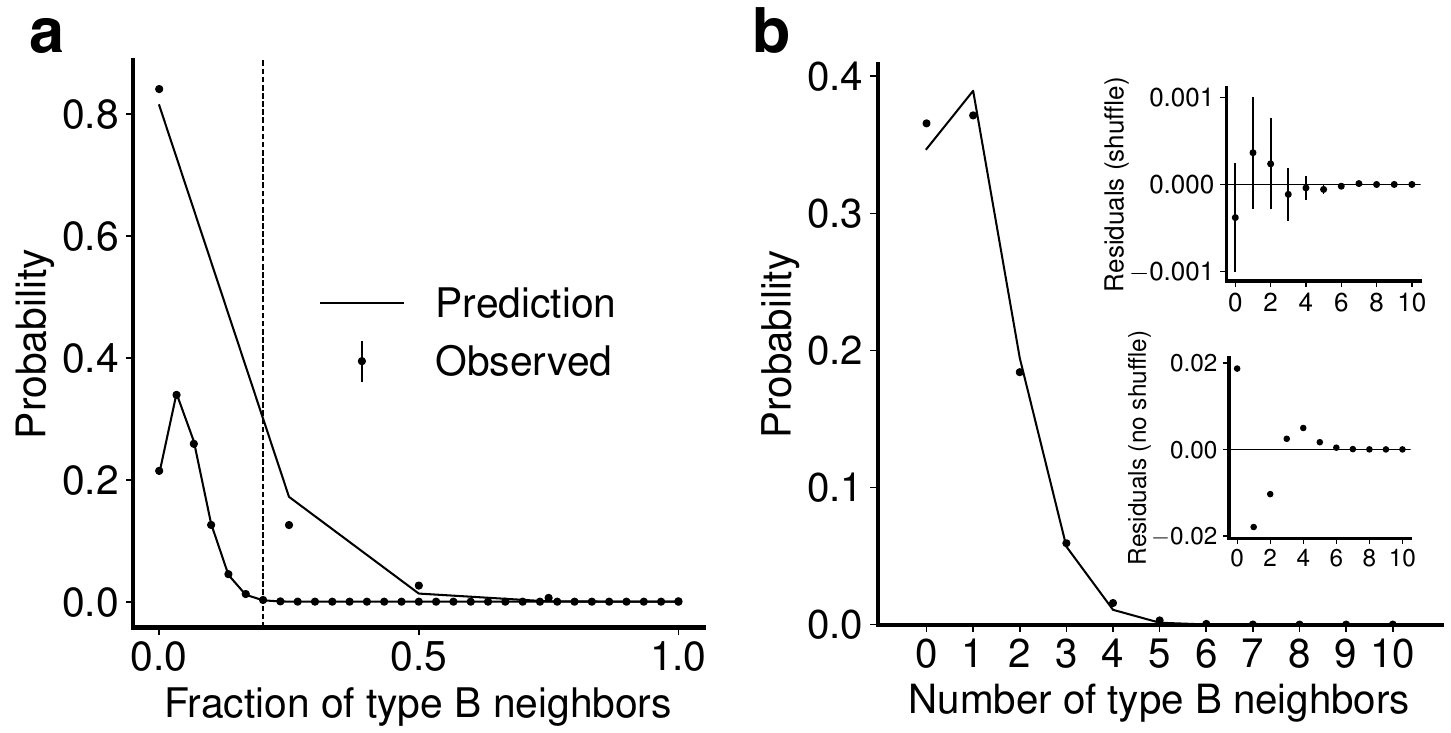}
 }
\caption{{\bf Local distribution of $y$ for regular random graphs.} The figure shows the expected distribution according to the assumption of ``no locality'' (see main text) as the black line, and compares with the observed distribution from simulations (dots) on random regular networks of size $N = 1000$. The error bars denote standard error. {\bf a,} Comparison of the distributions for $k = 4$ and $k = 30$, for a situation where $y_n = 0.2$, and the global value of $\bar{y} = 0.05$. Since $\bar{y} < y_n$, almost no individuals experience positive selection when $k$ is large and observed values of $y$ are tightly concentrated around the global value. However, for small $k$, a significant number of nodes do experience positive selection ``just by chance''. {\bf b,} Predicted and observed values on a network with $k = 10$ and $\bar{y} = 0.1$. The lower inset shows the residual mismatch between the prediction and the observed values, while the upper inset shows the residuals in simulations where the location of \ifc{} individuals is shuffled at every time step (making the no locality assumption exactly true by definition). Shuffling causes the mismatch to disappear. Note that locality slightly increases the chances of the extreme outcomes of having zero \ifc{} neighbors as well as the chances of having many \ifc{} neighbors. This is because \ifc{} nodes are created only if they are in contact with another \ifc{} individual, so they are slightly more likely than chance to be connected to each other in a locally ``tree-like'' structure \cite{dodds2011direct}. Because the distribution of $y$ is slightly wider than in our approximation, the variance is slightly higher and thus the effect on selection is slightly more positive than predicted. This explains the slight underestimation of $P_{reach}$ and $P_{fix}$ in {\bf Figure 3} in the main text.}
  \label{edfig:edfig2}
\end{figure}

\clearpage

\begin{figure}[!ht]
 \noindent\makebox[\linewidth]{
  \includegraphics[width=1.0\linewidth]{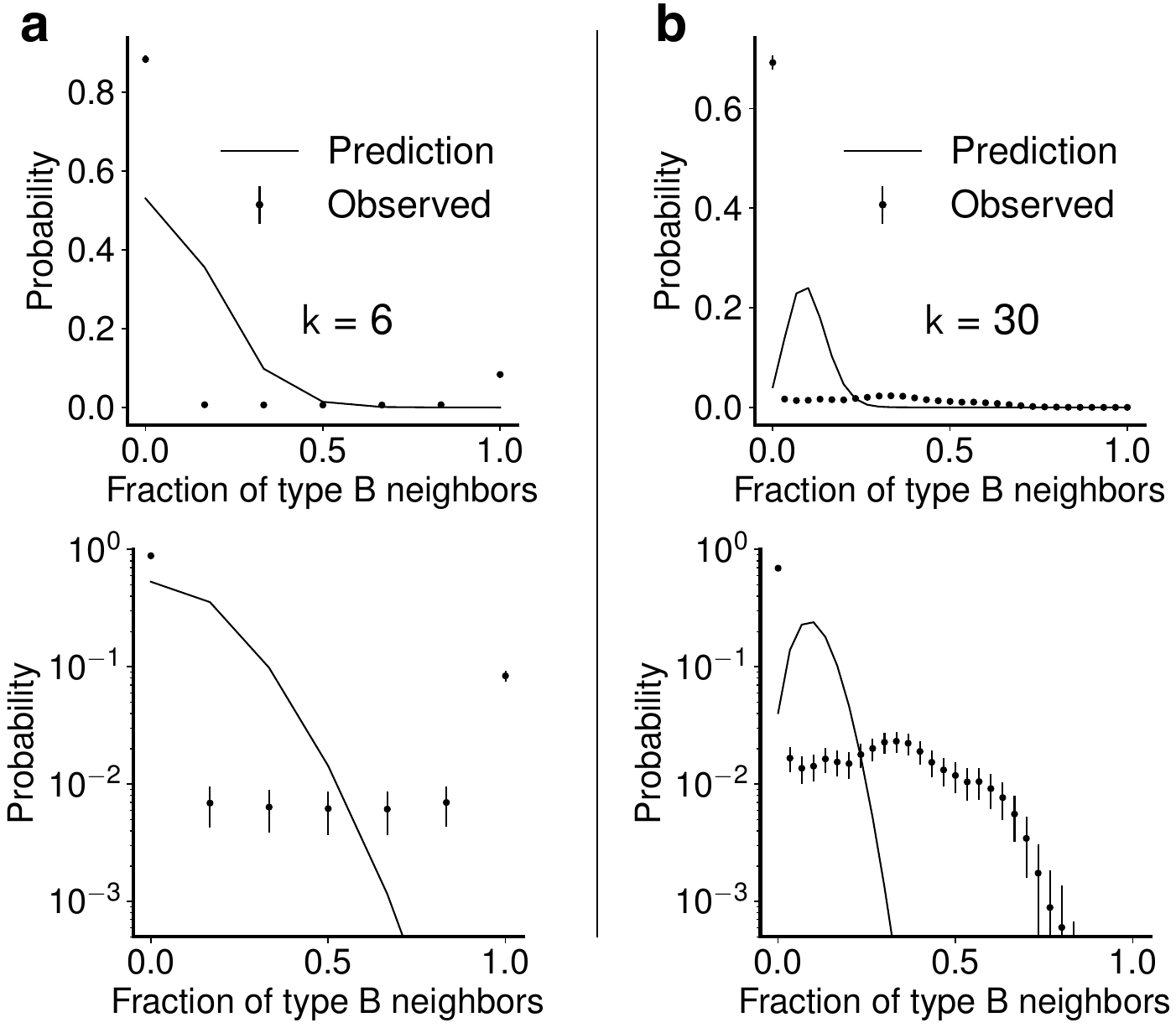}
 }
\caption{{\bf Local distribution of $y$ for 1D lattice networks.} The figure shows the expected distribution according to the ``no locality'' assumption for random regular graphs (black line, see main text), and the observed distribution (dots) for a contagion on a 1D lattice network with $\bar{y} = 0.1$, $N = 1000$ and $k = 6$ ({\bf a}) as well as $k = 30$ ({\bf b}). The error bars denote standard error. The bottom plots show the same data as the top, but on a logarithmic scale. It is clear that the ``locality'' on the lattice causes more extreme $y$ values that depart significantly from the ``no locality'' prediction. These patterns arise purely due to the network structure, even for simple contagions (here $\alpha = 0$ and $\beta = 0$).}
  \label{edfig:edfig3}
\end{figure}

\clearpage

\begin{figure}[!ht]
 \noindent\makebox[\linewidth]{
  \includegraphics[width=1.0\linewidth]{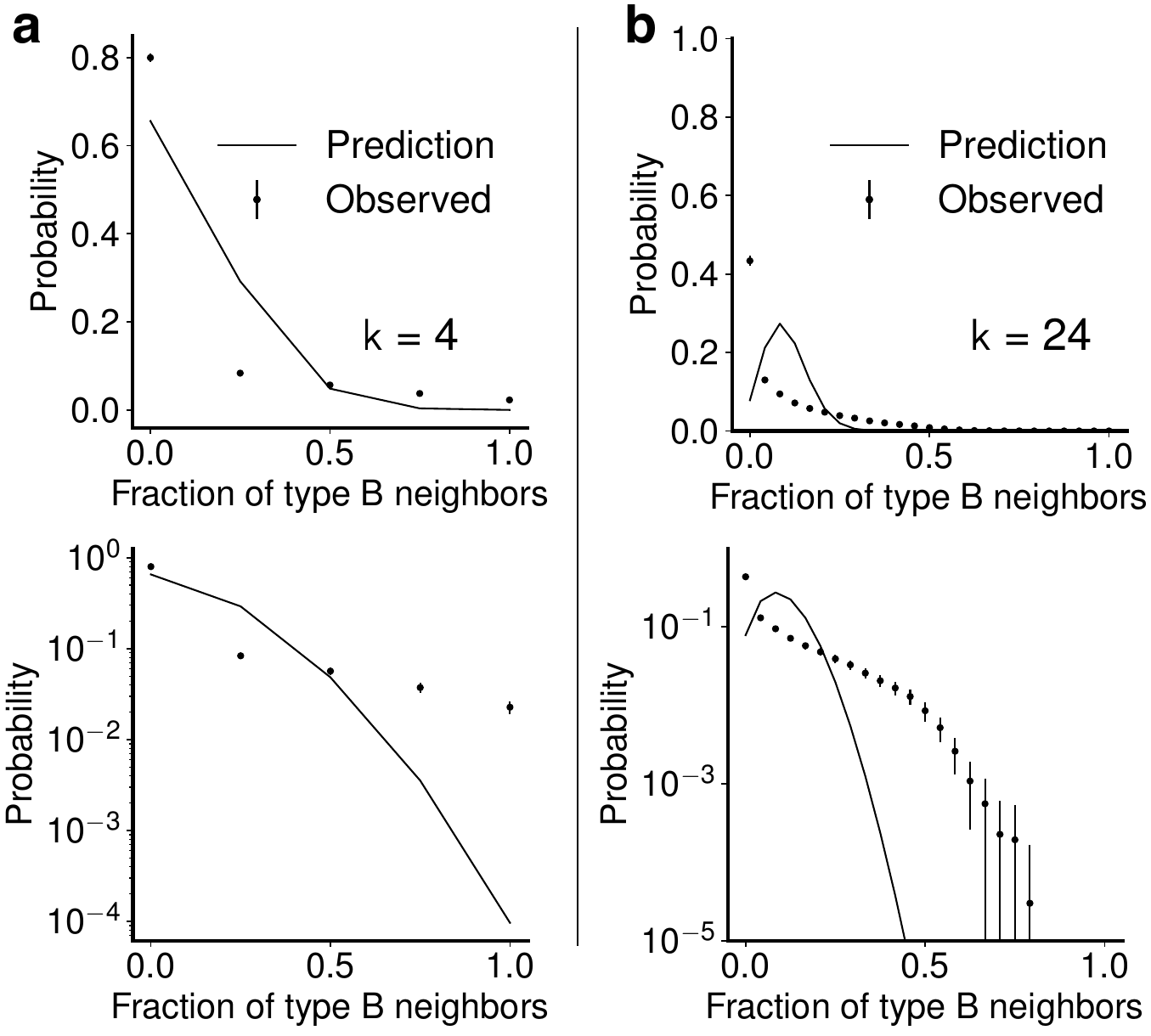}
 }
\caption{{\bf Local distribution of $y$ for 2D lattice networks.} The figure shows the expected distribution according to the ``no locality'' assumption for random regular graphs (black line, see main text), and the observed distribution (dots) for a contagion on a 2D lattice network with $\bar{y} = 0.1$, $N = 1600$ and $k = 4$ ({\bf a}) as well as $k = 24$ ({\bf b}). The error bars denote standard error. The bottom plots show the same data as the top, but on a logarithmic scale. It is clear that the ``locality'' on the lattice causes more extreme $y$ values that depart significantly from the ``no locality'' prediction. These patterns arise purely due to the network structure, even for simple contagions (here $\alpha = 0$ and $\beta = 0$).}
  \label{edfig:edfig4}
\end{figure}

\clearpage

\begin{figure}[!ht]
 \noindent\makebox[\linewidth]{
  \includegraphics[width=1.0\linewidth]{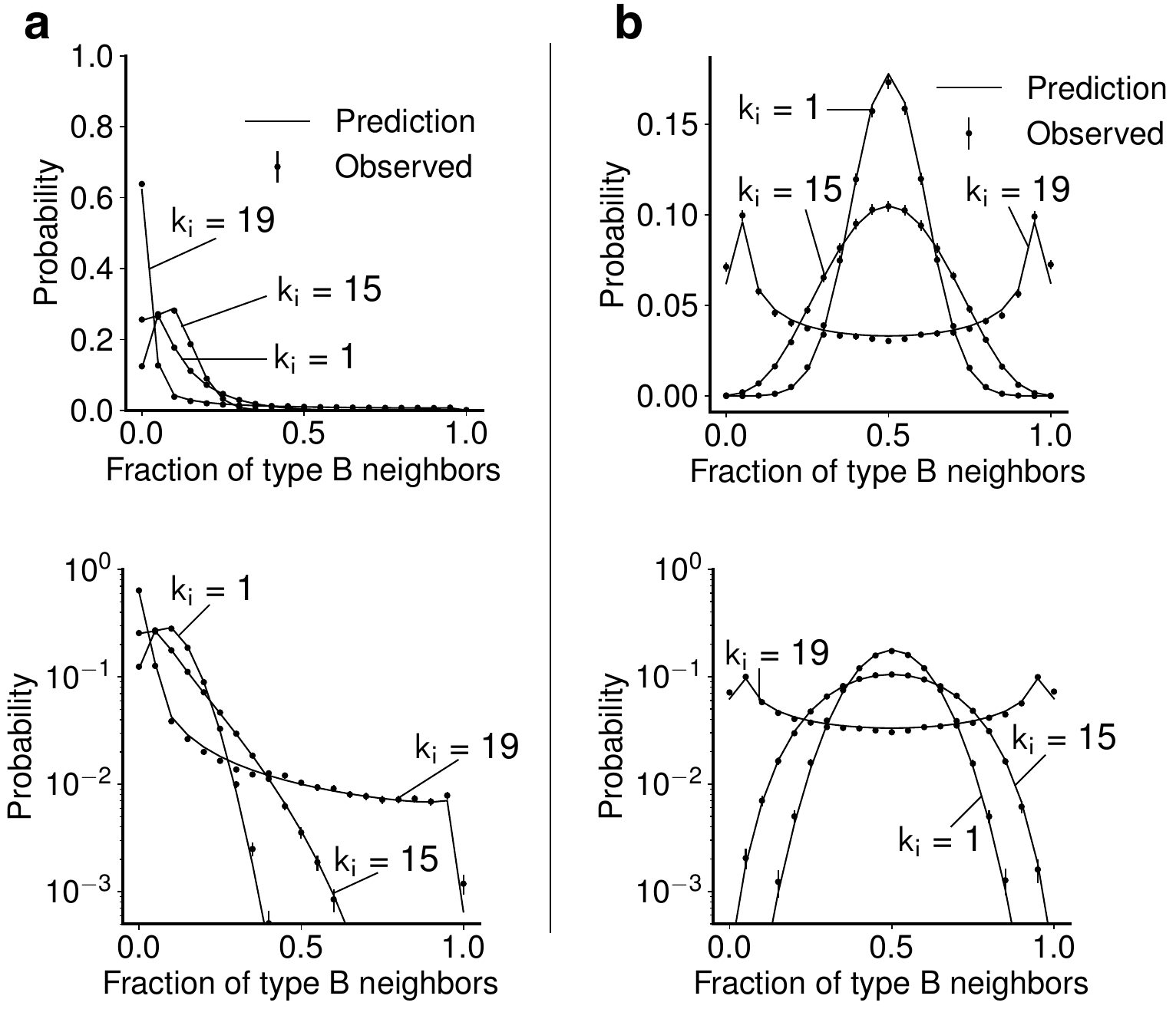}
 }
\caption{{\bf Local distribution of $y$ for community based networks.} The figure shows the expected distribution according to the equilibrium assumption (black line, see main text), and the observed distribution (dots) for a contagion on a network with $N = 1000$ and $k = 20$. We show distributions for $\bar{y} = 0.1$ ({\bf a}) and $\bar{y} = 0.5$ ({\bf b}). The error bars denote standard error. The bottom plots show the same data as the top, but on a logarithmic scale. The theoretical prediction matches well. When clustering and community strength reaches a critical value, the distribution shifts from a tightly concentrated one to a broad distribution with significant probability mass at the extremes. These patterns arise purely due to the network structure, even for simple contagions (here $\alpha = 0$ and $\beta = 0$).}
  \label{edfig:edfig5}
\end{figure}

\begin{figure*}[!ht]
 \noindent\makebox[\linewidth]{
  \includegraphics[width=1.0\linewidth]{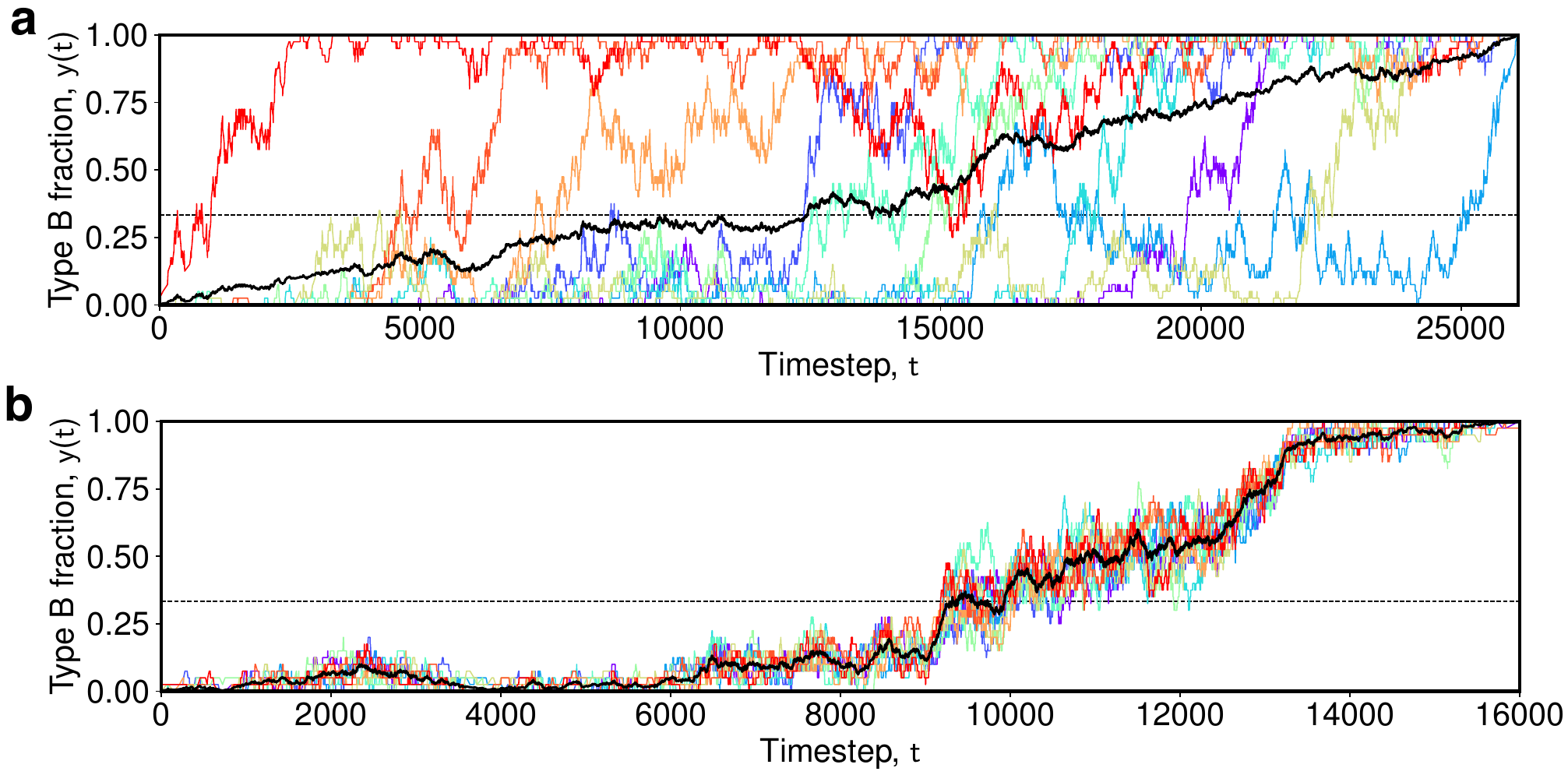}
 }
\caption{{\bf Contagion over time for community based networks.} The figure shows the temporal evolution of a contagion on a community based network. In the case of high community strength ({\bf a}), as in the high clustering case of the real network in {\bf Figure 1 d} in the main text, each community fixes one at a time. In the opposite case ({\bf b}), the $y$ values in each community are tightly coupled to the global value, as in {\bf Figure 1 e} in the main text. Parameters: $N = 400$, $y_n = \frac{\beta}{\alpha} = 0.2$ (dashed horizontal line), size of communities $m = 40$, $k_i = 39$ ({\bf a}) and $k_i = 1$ ({\bf b}).}
  \label{edfig:edfig6}
\end{figure*}

\clearpage

\begin{figure}[!ht]
 \noindent\makebox[\linewidth]{
  \includegraphics[width=1.0\linewidth]{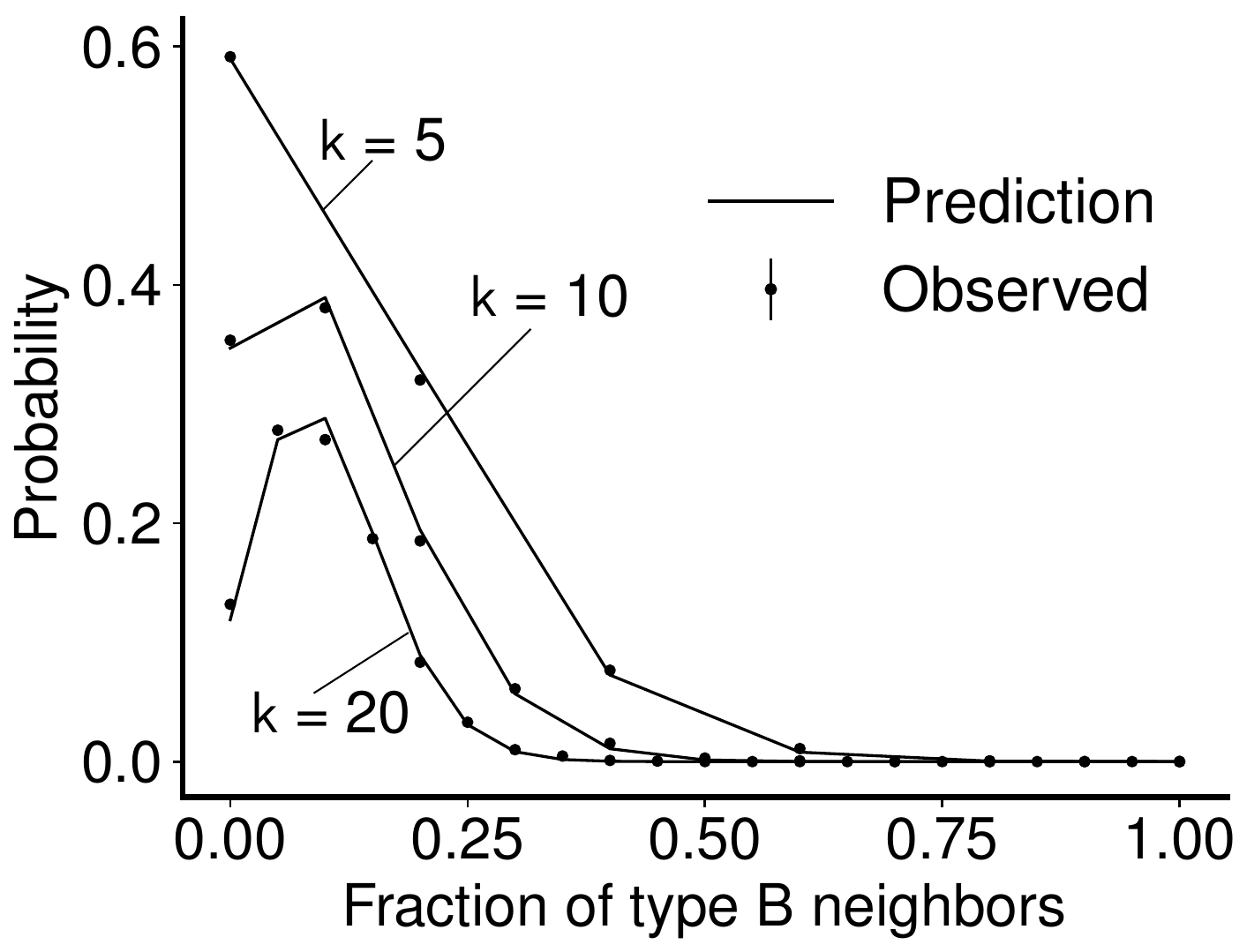}
 }
\caption{{\bf Local distribution of $y$ for networks with degree distributions.} The figure shows the expected distribution of $y$ as seen by nodes of degree $5$, $10$ and $20$ according to the ``no locality'' assumption (black line, see main text), as well as the observed distribution (dots), for a contagion on a network with $N = 1000$, mean degree $\bar{k} = 20$, $\bar{y} = 0.1$, and degree standard deviation $\sigma_k = 10$. Error bars denote standard error. The theoretical prediction matches well.}
  \label{edfig:edfig7}
\end{figure}

\clearpage

\begin{figure}[!ht]
 \noindent\makebox[\linewidth]{
  \includegraphics[width=1.0\linewidth]{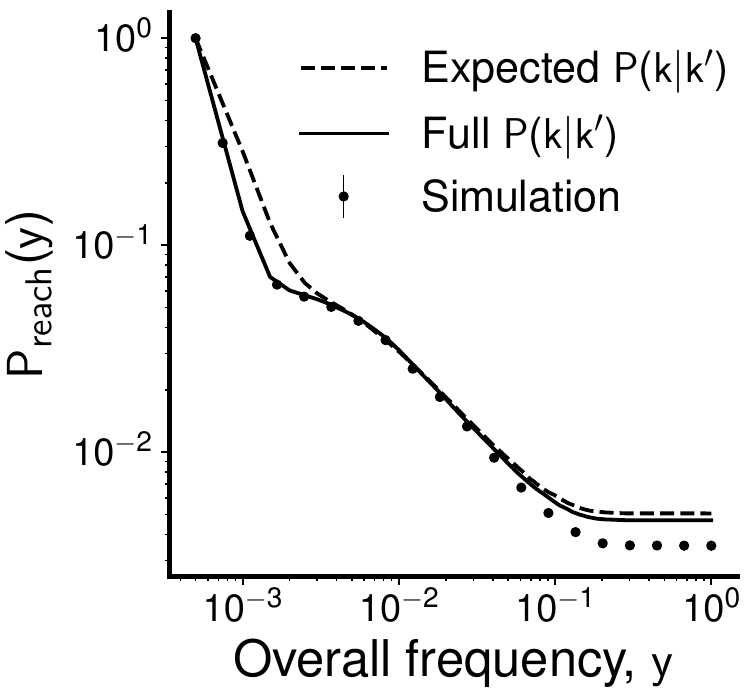}
 }
\caption{{\bf Effect of the neighbor degree distribution.} For simplicity we omit the bar for $\bar{y}$. The figure shows $P_{reach}(y)$ as computed according to the ``no locality'' assumption for a contagion on a network with $N = 10000$, mean degree $\bar{k} = 10$, degree standard deviation $\sigma_k = 30$, and strong positive degree correlations (see ``Networks with degree distrubtions'' for details). The lines show the prediction assuming the ``expected'' neighbor degree distribution $P(k | k') = \frac{k P(k)}{\bar{k}}$ (dashed), as well as the actual neighbor degree distribution $P(k | k')$ taking degree correlations into account (solid). The error bars denote standard error. Once the full neighbor degree distribution is taken into account, the results agree well with simulations (dots). We have verified that the residual mismatch at high $y$ disappears when \ifc{} nodes are shuffled at every time step (keeping their degree constant), showing the mismatch is due to the ``no locality'' assumption not being exactly true. Parameters: $\alpha = 1.0$, $\beta = 0.1$.}
  \label{edfig:edfig8}
\end{figure}

\clearpage

\bibliography{complexcontagions}